\documentclass[a4paper,10pt]{article}
\usepackage[utf8x]{inputenc}
\usepackage{threeparttable}
\pdfoutput=1
\usepackage{verbatim}
\usepackage{epsfig}
\usepackage{epsf}
\usepackage{epstopdf}
\usepackage{graphicx}
\usepackage{textcomp}
 \usepackage{amsmath}
\usepackage{amssymb}
\usepackage{color}
\definecolor{Blue}{rgb}{0.3,0.3,0.9}
\definecolor{red}{rgb}{1,0,0}

\newcommand{\bM}[1]{\mathbf{#1}}    

\title{\textbf{\fontsize{15}{15}\selectfont A Quantum Mechanical Review of Magnetic Resonance Imaging}}
\author{\smallskip \textbf{\fontsize{11}{11}\selectfont Stephen G. Odaibo$^{1,2}$}\\ \textbf{\fontsize{10}{10}\selectfont M.S.(Math), M.S.(Comp. Sci.), M.D.}\\ \newline\newline \small{\textsl{\fontsize{9}{9}\selectfont $^1$Quantum Lucid Research Laboratories, Clinical Applications and Theory Group}}\\ \small{\textsl{\fontsize{7}{7}\selectfont PO Box 2173}}\\\small{\textsl{\fontsize{7}{7}\selectfont Arlington, VA 22202}}\\\small{\color{blue}{\fontsize{7}{7}\selectfont stephen.odaibo@qlucid.com}}\\\newline\newline \\\small{\textsl{\fontsize{9}{9}\selectfont $^2$Howard University Hospital}}\\ \small{\textsl{\fontsize{7}{7}\selectfont Washington, D.C.  }} }
\date{}

\begin{document}

\maketitle

\begin{abstract}

In this paper, we review the quantum mechanics of magnetic resonance imaging (MRI). We traverse its hierarchy of scales from the spin and orbital angular momentum of subatomic particles to the ensemble magnetization of tissue. And we review a number of modalities used in the assessment of acute ischemic stroke and traumatic brain injury.

\end{abstract}
\hspace{-1.5mm}\indent\indent {{\fontsize{7}{7}\selectfont {\bf Keywords:} MRI, Quantum Mechanics, Magnetic Dipole, Stroke, TBI, Neuroimaging}}
\begin{center}
\line(1,0){300}
\end{center}

\section{Introduction}

\begin{figure}[ht]
\begin{center}
\scalebox{.65}
{\includegraphics{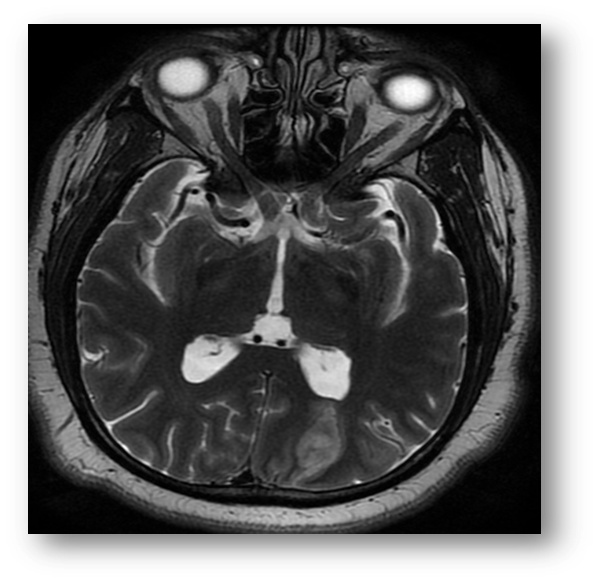}}
\end{center}
\caption{T2 axial fast spin echo image showing acute ischemia in the PCA distribution. There is relative hyperintensity in the left occipital region of infarction. Note the prominent hyperintensity of the vitreous and cerebrospinal fluid, both consistent with the hyperintensity of water on T2-weighted imaging.}
\label{fig:t2axfse}
\end{figure}

MRI plays a key role in the diagnosis and management of acute ischemic stroke and traumatic brain injury. A rigorous understanding of MRI begins with the elementary particles which give rise to the magnetic properties of tissue. Elementary particles have two independent properties which manifest magnetic moments: (i) orbital angular momentum, and (ii) intrinsic spin. These two properties are independent only to first order, as it was their very interaction via spin-orbit coupling which led to detectable energy differences in the hydrogen spectrum, and consequently to the discovery of intrinsic spin. We discuss both properties in this paper in the context of the hydrogen atom. The hydrogen atom $^1$H is the most frequently targeted nucleus in MRI due largely to its biological abundance and high gyromagnetic ratio. The $^1$H nucleus consists of a single proton. Protons are made of quarks, specifically one down and two up quarks, each of which are of spin 1/2. The charge and spin of the proton are both directly due to its quark composition. The down quark has a charge of -1/3 while the up quarks each have charge +2/3, which all sums up to a charge of +1. The nuclear spin is 1/2 by virtue of spin cancellation from the antiparallel alignment of two of the three quarks. 

\subsection*{Content Outline}

The remainder of this paper is organized as follows: Section~(\ref{sec:MRI}) presents an overview of the MRI, Section~(\ref{sec:stroke}) reviews the MRI assessment of acute ischemic stroke, Section~(\ref{sec:H_atom}) describes the hydrogen atom and reviews an analytical solution to its Schr\"{o}dinger equation, Section~(\ref{sec:spin}) reviews intrinsic spin, Section~(\ref{sec:adding}) reviews quantum mechanical addition of orbital angular momentum and spin, Section~(\ref{sec:groups}) reviews Group Theory of SU(2) and SO(3) and their relationship, and Section~(\ref{sec:summary}) summarizes this review. 

The MRI section, Section~(\ref{sec:MRI}), contains a number of subsections which are outlined as follows: Subsection~(\ref{subsec:scales}) discusses the hierarchy of scales in MRI phenomenology, Subsection~(\ref{subsec:magnetism_e_n}) reviews the magnetism of electrons and nucleons, Subsection~(\ref{subsec:mag_potential}) reviews the magnetic potential energy and its role in the descriptions of magnetization, torque, force, and larmor precession, Subsection~(\ref{subsec:pulse_mag}) reviews pulse magnetization, Subsection~(\ref{subsec:equilibrium_mag}) describes equilibrium magnetization, Subsection~(\ref{subsec:t1}) describes spin-lattice or T$_1$ recovery, Subsection~(\ref{subsec:t2}) describes spin-spin or T$_2$ decay, Subsection~(\ref{subsec:fid}) reviews free inductance decay, Subsection~(\ref{subsec:bloch}) reviews the Bloch equations, Subsection~(\ref{subsec:contrast}) overviews intravenous contrast agents in MRI, Subsection~(\ref{subsec:gradient}) reviews gradient-based band-selection method for MRI space localization, and Subsection~(\ref{subsec:machine}) reviews the components of the MRI machine.

The stroke section, Section~(\ref{sec:stroke}), contains a number of subsections which are outlined as follows: Subsection~(\ref{subsec:dwi}) reviews diffusion weighted imaging (DWI), Subsection~(\ref{subsec:pwi}) reviews perfusion weighted imaging (PWI), Subsection~(\ref{subsec:dwi_pwi}) reviews the combination of DWI and PWI and its utility in imaging the penumbra, Subsection~(\ref{subsec:mrs}) reviews magnetic resonance spectroscopy, Subsection~(\ref{subsec:bold}) reviews blood oxygen level dependent (BOLD) imaging, and Subsection~(\ref{subsec:mra}) reviews MRA.

Appnedix (A) reviews the signals processing involved in MRI, including the Fourier transforms, the canonical sampling theorem, and the convolution theorem.

\section{Magnetic Resonance Imaging}
\label{sec:MRI}

The MRI is an optical technology whose core equation is the Maxwell-Faraday Equation. It exploits ensemble phenomena in which the composition of a sample can be probed by sensing its magnetic properties through radio frequency waves. Here, we describe the various parts of MRI and their respective roles in the whole.



\subsection{A Hierarchy of Scales}
\label{subsec:scales}

The hierarchy of scales in the modeling of magnetic resonance imaging proceeds from the subnuclear scale of quarks and gluons, to the subatomic scale of discrete electrons and nucleons, and finally to the bulk sample scale where ensemble effects and statistical mechanics are the operative physics. In a general sense, the subnuclear scale is coupled to quantum field theory, the subatomic scale is coupled to quantum mechanics, and the bulk sample scale is coupled to classical electromagnetics. 

Certain phenomena in MRI are only meaningful over specific regimes. For instance, a magnetic moment experiences a torque when an external magnetic field is applied. However, the concept of torque is a classical concept. It is deterministic and continuous, and is meaningful only at the level of ensemble or bulk effect. On the microscopic level, such as on the scale of single electrons and nucleons, quantum mechanics is the operative physics, and state transitions are quantized. This is encoded in the time evolution of quantum states governed by an appropriate Hamiltonian, yielding the specific Schr\"{o}dinger equation that applies in that setting. 

Although pedagogically separable, the scales are intrinsically linked physically. For instance, the bulk magnetization, $\bM{M}$, is the net sum of the atomic level magnetic moments, $\boldsymbol{\mu}_j$. Similarly the pulse magnetization frequency used to torque the bulk magnetization vector and change its direction is a radio frequency wave, $\bM{B}_1$, oscillating at the larmor frequency of the individual nucleons. Furthermore, $\bM{B}_1$ takes its effect by acting directly on the individual electrons and nucleons, causing them to precess in phase.

\subsection{Magnetism of Electrons and Nucleons}
\label{subsec:magnetism_e_n}

 Electrons, protons, and neutrons are magnets. They derive their magnetism from intrinsic spin and orbital angular momentum. In the presence of an externally applied magnetic field, this results in bulk magnetism of sample tissue. The magnetic dipole moment, $\boldsymbol{\mu}$, of a single unpaired electron or nucleon is given by,

\begin{equation}
\mbox{\textbf{\textmu}}= \gamma\bM{S}, 
\end{equation}

where $\gamma$ is the gyromagnetic ratio measured in Hz/T (Hertz per Tesla) and $\bM{S}$ is the spin. The gyromagnetic ratio for the electron is given by,

\begin{equation}
 {\gamma}_e = g_e \frac{e}{2m_e},
\end{equation}

where $e$ is the elementary charge, $m_e$ is the mass of the electron, and $g_e$ is the electron gyromagnetic factor (g-factor), a dimensionless number whose experimental agreement with theory is near unprecedented on this scale, and thereby lends strong credibility to the theory of quantum electrodynamics. Table~(\ref{Tab:consts}) shows the electron and proton g-factors. The gyromagnetic ratio for nucleons is given by,

\begin{equation}
 {\gamma}_n = g_p \frac{e}{2m_p},
\end{equation}

where $m_p$ is the mass of the proton, and $g_n$ is the g-factor of the nucleon. Estimated gyromagnetic ratios of some biologically-relevant nuclei are shown in Table~(\ref{Tab:Gyromag})~\cite{beki2004,li2012}.

\begin{table}[ht]
 \caption{Gyromagnetic ratios of some biologically-relevant nuclei}
\centering
\small{
\begin{tabular}{l|l|c|c}
\hline\hline
Element & Nuclei &$\gamma_n~(10^6~\mbox{rad}~\mbox{s}^{-1}~\mbox{T}^{-1} )$ & $\gamma_n/2\pi$(MHz T$^{-1}$) \\
\hline\hline
Hydrogen & $^1$H & 267.513&42.576 \\
Deuterium & D, $^2$H & 41.065&6.536 \\
Helium & $^3$He & -203.789&-32.434 \\
Lithium & $^7$Li & 103.962&16.546 \\
Carbon  & $^{13}$C & 67.262& 10.705\\
Nitrogen  & $^{14}$N & 19.331& 3.007\\
Nitrogen  & $^{15}$N & -27.116& -4.316\\
Oxygen  & $^{17}$O &-36.264 &-5.772 \\
Fluorine  & $^{19}$F & 251.662& 40.053\\
Nitrogen  & $^{14}$N & 19.331& 3.007\\
Sodium  & $^{23}$Na & 70.761& 11.262\\
Phosphorus  & $^{31}$P & 108.291& 17.235\\
\hline
\end{tabular}
}
 \label{Tab:Gyromag}
\end{table}

The Bohr magneton, ${\mu}_B$, and the nuclear magneton, ${\mu}_N$, are named quantities related to the electron and proton gyromagnetic ratios respectively, and are given by,

\begin{equation}
 {\mu}_B = \frac{e\hbar}{2m_e},
\end{equation}
and
\begin{equation}
 {\mu}_N = \frac{e\hbar}{2m_p}.
\end{equation}

The magnetic moment of the electron is much larger than that of the nucleons, and as is discussed below, this corresponds to smaller precession, pulse, and signal frequency for protons than electrons. Specifically, nucleons precess, absorb, and emit electromagnetic signals in the radio frequency range. This range is non-ionizing radiation, and makes MRI a safer choice than ionizing imaging modalities such as plane and computed tomography X-rays.

\begin{figure}[ht]
\begin{center}
\scalebox{.50}
{\includegraphics{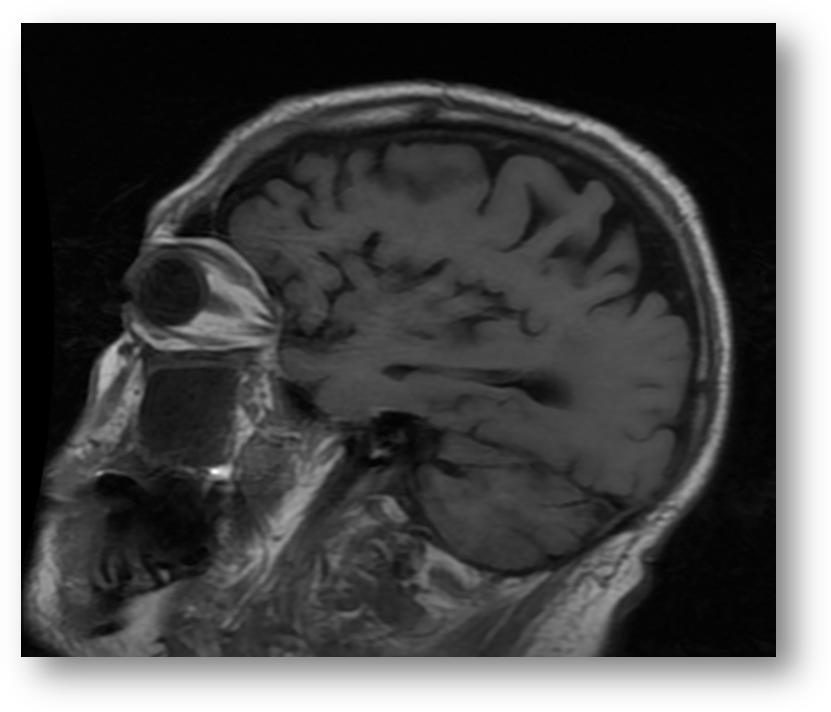}}
\end{center}
\caption{T1 FLAIR. Sagittal section. Note the hyperintensity of orbital fat, and the hypointensity of vitreous and cerebrospinal fluid. Hyperintensity of fat and hypointensity of water are radiologic characteristics of T1 weighted imaging.}
\label{fig:sag_t1_flair}
\end{figure}

\subsection{The Magnetic Potential Energy}
\label{subsec:mag_potential}

The magnetic potential energy is, 

\begin{equation}
 U = -\boldsymbol{\mu} \cdot \bM{B}.
\end{equation}

The force on a magnetic moment, the torque on a magnetic moment, and the larmor precession frequency or resonance frequency, can all be described in terms of the magnetic potential energy, $U(\bM{r},t)$. 

\subsubsection{Magnetization}

The magnetic moment aligns itself in either parallel or anti-parallel orientation to the external field $\bM{B}_0$. The extent of parallelism is constrained by the eigenvalues of the $\bM{S}^2$ and $\bM{S}_z$ operators above, such that from the geometry, the parallel and anti-parallel states make angles of $\cos^{-1}(1/\sqrt{3}) \approx 54.7356103172$ degrees with the positive and negative z-axes respectively (assuming $\bM{B}_0$ is oriented exclusively along the z-axis).

This process is called magnetization. In the case of a population sample, upon application of $\bM{B}_0$, the spin orientation distribution goes from a somewhat arbitrary array of directions to being composed of only two directions, parallel and antiparallel. The parallel orientation is the lower energy state, and thereby is the more populous state for protons in the sample. The population distribution is temperature dependent, such that the proportion of protons in the higher energy state increases with temperature. The relationship is given by,

\begin{equation}
 \frac{N_{h}}{N_{l}} = e^{-E/kT},
\label{boltz}
\end{equation}

where $N_{h}$ is the higher energy state, $N_{l}$ is the lower energy state, $E$ is the energy difference between the two states, $k$ is the Boltzmann constant, and T is the absolute temperature in Kelvin. The bulk magnetization can be modeled as,

\begin{equation}
 \bM{M}(t) = \sum_j{\boldsymbol{\mu}_j(t)}.
\end{equation}

with ${\bM{B}_0}=\hat{\bM{z}}B_0$, it follows that $M(0)=M_z(0)$ and $M_{xy} = 0$. In other words, the equilibrium magnetization is entirely along the $z$ (longitudinal) direction and there is no component in the $xy$ (transverse) direction. 

\subsubsection{Torque}
The torque on a magnetic moment in a magnetic field acts to align it with the field, is a function of the magnetic potential energy, and is given by,

\begin{equation}
 \tau(\alpha) = -\frac{\partial}{\partial\alpha}(-\boldsymbol{\mu}\cdot\boldsymbol{B}),
\end{equation}

where $\alpha$ is the angle between $\bM{B}$ and $\boldsymbol{\mu}$, and the minus sign indicates the restoring nature of the torque.

\subsubsection{Force}

The force on a nucleon or electron in a magnetic field is,

\begin{equation}
 \bM{F} = -\nabla(-\boldsymbol{\mu}\cdot\bM{B} ).
\end{equation}

\subsubsection{Precession}

When an external magnetic field $\bM{B}_0$ is applied to an electron or neutron particle, the particle's magnetic moment precesses about the direction of $\bM{B}_0$at a frequency called the larmor frequency given by,

\begin{equation}
 \omega = \gamma B_0.
\end{equation}

In the equilibrium magnetization state with ${\bM{B}_0}=\hat{\bM{z}}B_0$ and  $M_{xy} = 0$, there is no precession of the bulk magnetization $\bM{M}$ because $\boldsymbol{\mu} = \bM{M}\times {\bM{B}}_0 = 0$.

Considering only the spin of a spin 1/2 object in a magnetic field, and assuming $B = {\bM{\hat{e}}}_z$, we get the following Hamiltonian,

\begin{equation}
 H = -\boldsymbol{\mu}\cdot\bM{B} = -(\frac{-g_xe}{2m_x}\bM{S})\cdot\bM{B}:= \boldsymbol{\omega}\cdot\bM{S},
\end{equation}

where the subscript $x$ denotes either $e$ for electron or $p$ for a neutron, and where,

\begin{equation}
 \boldsymbol{\omega} :=  \frac{g_xeB}{2m_e}{\bM{\hat{e}}}_z.
\end{equation}

It follows that,

\begin{equation}
 H = \omega S_z.
\end{equation}

We see that remarkably, the Hamiltonian is proportional to the z-component of spin, therefore the magnetic spin eigenstates are simultaneously energy eigenstates. Furthermore the energy eigenvalues are proportional to the magnetic spin quantum numbers. Specifically,

\begin{equation}
 E_{\pm} = \pm \frac{1}{2}\hbar\omega.
\end{equation}

The above relation led early researchers to the discovery of spin, by its manifestation as the Zeeman effect and the splitting of energy levels in the hydrogen optical spectrum.

\subsection{Pulse Magnetization}
\label{subsec:pulse_mag}

Given a bulk magnetization $\bM{M}$ due to a static magnetic field $\bM{B}_0=B_0\bM{\hat{z}}$, a pulse of a weaker magnetic field, $\bM{B}_1$, oscillating at the larmor frequency and applied perpendicular to $\bM{B}_0$, will result in a transition of the individual nucleons from dephased to in-phase oscillation. This manifests as the acquisition of a transverse phase $\bM{M}_{xy}$ and oscillations of $\bM{M}$ about the z-axis at the larmor frequency $\omega$. This ``tilting'' of the magnetization vector also implies a decrease in $M_z$. During the pulse, the net external magnetic field is the sum of $\bM{B}_0$ and $\bM{B}_1$. If the pulse duration was extended indefinitely, the result would be alignment of the bulk magnetization with the new direction, $\bM{B}_0+\bM{B}_1$, and there would be no resultant oscillation observed. The pulsed nature of $\bM{B}_1$ is therefore critical to the technology of magnetic resonance imaging.

\subsection{Equilibrium Magnetization}
\label{subsec:equilibrium_mag}

The net magnetization is given by,

\begin{equation}
 \bM{M} = (N_l-N_h)\boldsymbol{\mu},
\end{equation}

while the ratio of $N_l$ to $N_h$ is given in Equation~(\ref{boltz}) above as,  $\frac{N_{h}}{N_{l}} = e^{-E/kT}$. At equilibrium, the solution of the above system of two equations yields the equilibrium magnetization,

\begin{equation}
 M_0 = \frac{B_0\gamma^2\hbar^2}{4\mbox{k\hspace{-.5em}-} T}\rho_p,
\end{equation}
 
where $\mbox{k\hspace{-.5em}-}=k/2\pi$ is the reduced Boltzmann constant and $\rho_p$ is the proton density.

\subsection{Spin-Lattice Relaxation (T1 Relaxation)}
\label{subsec:t1}

The spin-lattice relaxation, also known as $T_1$ relaxation or longitudinal relaxation, is the process of longitudinal magnetization recovery following a $\bM{B}_1$ perturbation. Specifically, $T_1$ is the time constant of longitudinal magnetization recovery, and is given by,

\begin{equation}
 M_z(t) = M_0(1-e^{-t/T1}) + M_z(0)e^{-t/T1},
\label{Eqn:T2}
\end{equation}

where $M_0$ is equilibrium magnetization and $M_z(0)$ is the longitudinal magnetization instantaneously after pulse excitation. In inversion recovery, the initial magnetization is negative, and typically $M_z(0)=-M_0$. Substituting this value into Equation~(\ref{Eqn:T2}) above yields,

\begin{equation}
  M_z(t) = M_0(1-2e^{-t/T1}) 
\label{Eqn:InvRecT2}.
\end{equation}

\begin{figure}[ht]
\begin{center}
\scalebox{.70}
{\includegraphics{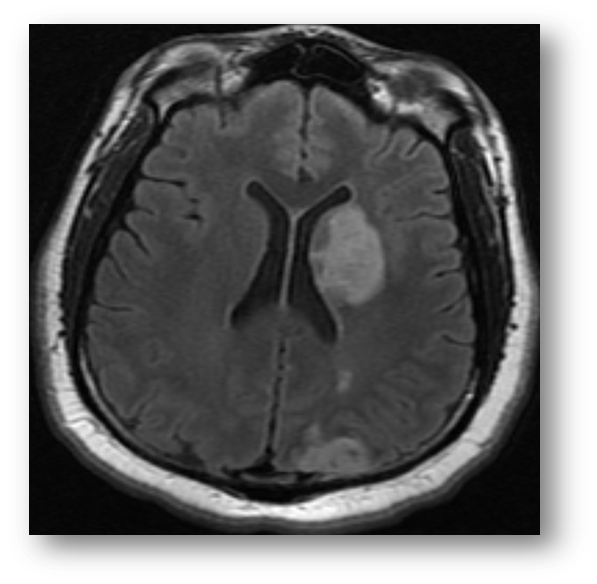}}
\end{center}
\caption{T2 FLAIR showing hyperintense signal in the PCA and MCA distributions, consistent with acute ischemic stroke. Note that the hypointensity of cerebrospinal fluid due to fluid-attenuation allows for better contrast with the periventricular lesion.}
\label{fig:t2_flair_b}
\end{figure}

Fluid attenuated inversion recovery (FLAIR) is useful in the diagnosis of certain pathologies such as periventricular white-matter lesion in multiple sclerosis. FLAIR can be T1 or T2 weighted. Figure~(\ref{fig:sag_t1_flair}) shows a sagittal section of a T1 FLAIR brain image, and Figure~(\ref{fig:t2_flair_b}) shows a T2 FLAIR brain image.

\subsection{Spin-Spin Relaxation (T2 Relaxation)}
\label{subsec:t2}

The spin-spin relaxation, also known as $T_2$ relaxation or transverse relaxation, is the process of transverse magnetization relaxation following a $\bM{B}_1$ perturbation. Specifically, $T_2$ is the time constant of transverse magnetization recovery, and is given by,

\begin{equation}
 M_{xy}(t) = M_{xy}(0)e^{-t/T2},
\end{equation}

where $M_{xy}(0)$ is the transverse magnetization instantaneously after pulse excitation. Following the $B_1$ induction of in-phase precession, the proton spins again begin to dephase according to $T_2$ time constant. Of note, the relaxation of transverse magnetization is faster than the recovery of longitudinal magnetization. In other terms, $T_2<T_1$.  Figure~(\ref{fig:t2axfse}) shows a T2 fast spin echo image of acute ischemic stroke in the posterior circulation.

\subsection{Free Inductance Decay}
\label{subsec:fid}

Upon pulse excitation and tilting, the magnetization vector oscillates about the z-axis. This oscillating magnetic moment in turn generates an electromotive force in accordance with Faraday-Maxwell law of electromagnetic induction,

\begin{equation}
 \nabla \times \bM{E} = -\frac{\partial \bM{B}}{\partial t},
\end{equation}

where $\nabla\times$ is the curl operator, $\bM{E}(\bM{r},t)$ is the electric field, and $\bM{B}(\bM{r},t)$ is the magnetic field. The above equation can be rendered in integral form via Stokes theorem to yield,

\begin{equation}
\oint_{\partial \Sigma} \bM{E}\cdot d\bM{s} = -\frac{\partial}{\partial t}\iint_{\Sigma} \bM{B}\cdot d\bM{A},
\end{equation}

where $\partial \Sigma$ can represent a wire coil bounding a surface $\Sigma$, $d\bM{s}$ is an infinitesimal segment length along the wire, and $d\bM{A}$ is vector normal to an infinitesimal area of $\Sigma$. Such integral transformation has certain advantages in numerical computation. For instance, Green's functions approaches can be invoked and can greatly decrease the number of unknowns, thereby decreasing the computational resource demand of the problem.

The radio frequency (RF) coils within the MRI probe experience an electromotive force due to the rotating magnetization. This results in an electric current signal which is fed as output to a screen. The signal decays with time according to the $T_2$ time constant, hence the name Free Inductance Decay (FID) signal~\cite{st1967}.

\subsection{Bloch Equations}
\label{subsec:bloch}

The Bloch equations describe the time evolution of the bulk magnetization vector, and are therefore also referred to as the equations of motion of magnetization. They are given by,

\begin{equation}
 \frac{\partial M_x(t)}{\partial t} = \gamma (\bM{M}(t)\times\bM{B}(t))_x - \frac{M_x(t)}{T_2},
\end{equation}

\begin{equation}
 \frac{\partial M_y(t)}{\partial t} = \gamma (\bM{M}(t)\times\bM{B}(t))_y - \frac{M_y(t)}{T_2},
\end{equation}
and
\begin{equation}
 \frac{\partial M_z(t)}{\partial t} = \gamma (\bM{M}(t)\times\bM{B}(t))_z - \frac{M_z(t)-M_0}{T_1},
\end{equation}

where $\gamma$ is the gyromagnetic ratio of the nucleon and $\bM{B}(t) = \bM{B}_0 + \bM{B}_1(t)$

\subsection{Intravenous MR Contrast Agents}
\label{subsec:contrast}

There are several contrast agents in current clinical use, most of which are based either on low molecular weight chelates of gadolinium ion (Gd$^{3+}$)~\cite{cael1999,br2002} such as Gadolinium-DPTA, or are based on iron oxide (FeO) such as Fe$_3$O$_4$, Fe$_2$O$_3$, and $\gamma$-Fe$_2$O$_3$ (or gamma phase maghemite)~\cite{bukr2004, baho2012, lech2012, juja1995}. Based on their magnetic properties, they confer contrast of vasculature from background, and by proxy of vascularization density, of organs from surround. Specifically, the Gd$^{3+}$-based agents predominantly shorten $T_1$ time, while the FeO-based agents predominantly shorten $T_2$ time. Nanoparticle formulations of contrast agents are in various stages of development and can be based on gadolinium or iron-oxide, or on other agents such as cobalt, nickel, manganese, or copper ions~\cite{tulo2012, lats2012, baho2012}. Optimization of the efficacy of MR contrast agents is an active area of research~\cite{ca2006,kabu1988,kobr2003,rapi2005,mika2001,jawy1990,koar2001}. The growing array of MR contrast applications has included efforts to target tissue or tumor-specific receptors~\cite{koar2001,liam2006,ca2009,wiko1997,faku1995,moth2003,stsh2012}. Serious adverse reactions to intravenous MRI contrast agents are rare. However in the mid 2000s gadolinium chelates were increasingly hypothesized as playing a role in nephrogenic systemic fibrosis (NSF), a disease entity unknown prior to 1997. It has since been essentially established that Gadodiamide, a gadolinium-containing agent, is associated with an increased risk of developing NSF in susceptible individuals ~\cite{bequ2012, kuka2007, ryth2008, th2006, mask2006}. Superparamagnetic iron oxides have been promoted by some as possible alternatives to gadolinium-based agents in patients at risk for NSF~\cite{neha2008}.

\subsection{Gradient-based 3D Spatial Localization}
\label{subsec:gradient}

For 3D localization, a gradient coil generates a magnetic field gradient in the bore. Then in accordance with the larmor frequency relation $\omega = \gamma B(r,t)$, it follows that the resonance frequency, $\omega = \omega(r,t)$ acquires space dependence, and thereby serves to effectively label the tissue in space. The process of space-specific excitation of the sample is called \textit{slice selection}. The excitation pulse frequency is simply chosen as a bandwidth frequency flanking the region-of-interest or the \textit{slice}. The process of \textit{scanning} is an iteration, however simple or complex, through a sequence of slices. And slice thickness is proportional to excitation pulse bandwidth. The slice thickness can be decreased by decreasing the bandwidth, or alternatively by increasing the magnetic field gradient. For instance, given a constant magnetic field gradient, the slice thickness between two points $r_1$ and $r_2$ is given by,

\begin{equation}
 \frac{\omega(r_2)-\omega(r_1)}{\gamma [B(r_2)-B(r_1)]} = r_2 - r_1
\end{equation}

Both the bandwidth and magnetic field gradient are Graphical User Interface (GUI)- adjustable parameters on current day MRI machines. This gradient-based localization technique was Paul Lauterbur's contribution to the development of MRI~\cite{la1973}.

\subsection{MRI Machinery}
\label{subsec:machine}

The MRI is a simple machine in principle. It was was invented by Raymond V. Damadian by 1971~\cite{da1971,da1974,da1973, wava2012}. It consists of a magnet, a radio frequency coil, an empty space or bore, and a computer processor. The magnet is for applying the static and pulsed magnetic fields. Current day machines typically use superconducting electromagnets. These are wire coils cooled to low temperatures using liquid helium and liquid nitrogen ~\cite{cuvi1974, lich1970, biha1972, hu1971, haei1972, thdu1971, me1971, wava2012}, allowing for greatly decreased electrical resistance, and consequently high current loops generating strong magnetic fields. The RF coil typically serves both as the generator of the pulsed magnetic field, and as the detector of the FID signal. Such duality of function is possible simply because the Maxwell-Faraday equation is true in both directions. Several machines also have a separate gradient coil. The gradient coil has $x$, $y$, and $z$ components indicating the spatial direction of the gradient field, and together conferring 3D localization. The computer processor runs the software instructions for pulse sequences and signals processing, including image reconstruction.

\subsubsection{Superconductivity}
MRI requires strong magnetic fields. Though the Earth's magnetic field is variable with time and is asymmetric between northern and southern hemispheres~\cite{opme2004}, it is estimated to range between 25,000 and 65,000 nanoteslas (0.25 to 0.65 Gauss) at the surface, with a commonly cited average of 46,000 nanotesla (0.46 Gauss)~\cite{wiwi1972}. Current clinical use MRIs commonly range from 1.5 Tesla to 7 Teslas. Therefore MRIs may require magnetic fields up to 280,000 times the earth's magnetic field. This has been achieved by a shift from permanent magnets to superconducting electromagnets, and presents another interdisciplinary link to both theory and practice. The dominant physical theory of superconductivity is called BCS theory~\cite{ba1955,co1956,baco1957,baco1957b} after its formulators, Bardeen, Cooper, and Schrieffer, and is based on electron pairing via phonon-exchange to form so called Cooper pairs. Though electrons are fermions and subject to the Fermi exclusion principle, precluding any two electrons occupying the same state, Cooper pairs act like bosons because their pairing and spin summation results in integer-spin character, a type of boson-like state. Upon cooling, these boson-like pairs can condense, analogous to Bose-Einstein condensation and representing a quantum phase transition. Quantum phase transitions have historically been challenging to study in a controlled way in the laboratory, but have recently been demonstrated by H. T. Mebrahtu et. al. in tunable quantum tunnelling studies of Luttinger liquids~\cite{mebo2012}. In superconductivity, upon transition into the condensed state, the coherence of the system imposes a high energy penalty on resistance, effectively resulting in a zero resistance state. This is formally encoded in the theory by an energy gap at low (subcritical) temperatures. In superconducting electromagnets, the coil cooling is done through liquid helium and liquid nitrogen and presents a connection to the chemistry of phase transitions, condensed matter and solid state physics, as well as the engineering aspects of attaining and maintaining such low temperatures.

\section{MRI Assessment of Acute Ischemic Stroke}
\label{sec:stroke}

In this section, we review the MRI modalities used in the assessment of acute ischemic stroke. 

\begin{figure}[ht]
\begin{center}
\scalebox{.75}
{\includegraphics{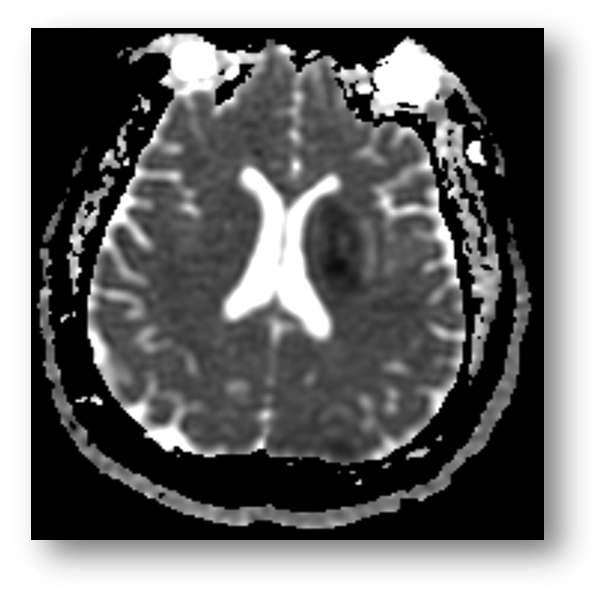}}
\end{center}
\caption{ADC map showing hypointensity in the left periventricular MCA distributions, consistent with acute ischemic stroke. Note the marked hyperintensity of the vitreous and cerebrospinal fluid. Both are fluid-filled cavities and have a much higher diffusion coefficient than tissue.}
\label{fig:adc_stroke_1}
\end{figure}

\begin{figure}[ht]
\begin{center}
\scalebox{.70}
{\includegraphics{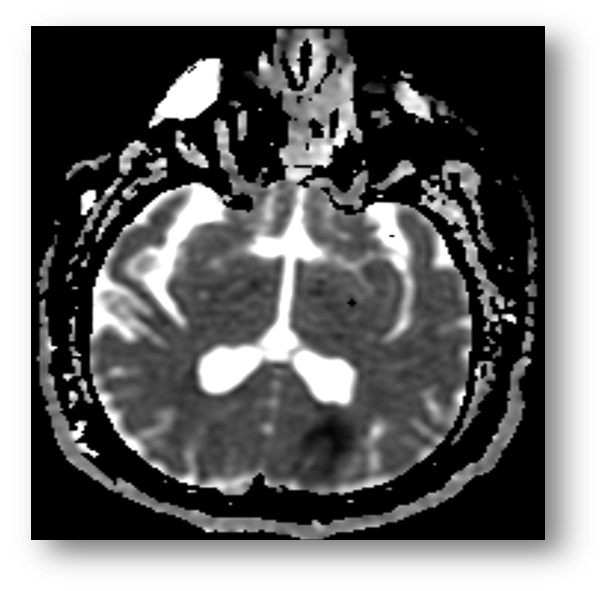}}
\end{center}
\caption{ADC map showing hypointensity in the PCA distribution, consistent with a parieto-occipital acute ischemic stroke}
\label{fig:adc_pca_stroke_b}
\end{figure}

\begin{figure}[ht]
\begin{center}
\scalebox{1.0}
{\includegraphics{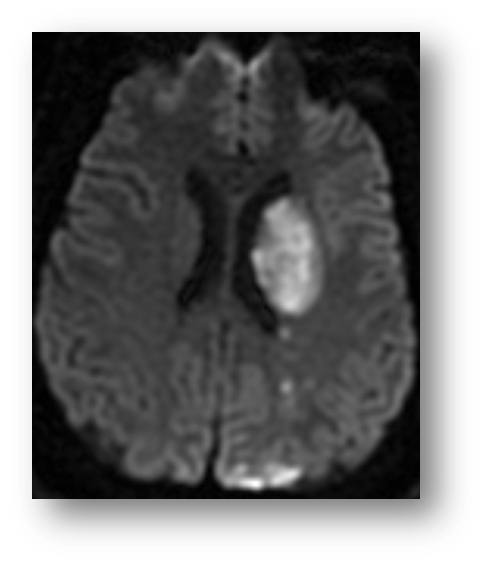}}
\end{center}
\caption{DWI showing hyperintensity in the left PCA and MCA distributions, consistent with acute ischemic stroke.}
\label{fig:dwi_2}
\end{figure}

\subsection{Diffusion-Weighted Imaging}
\label{subsec:dwi}

The diffusion-weighted MRI (DWI) modality exploits the Brownian motion of water molecules within a sample, and the relative phase shift in moving water versus stationary water. The Apparent Diffusion Coefficient (ADC) is a parameter which can be mapped to provide diagnostic information. In acute ischemic stroke, cytotoxic cellular injury results in axonal edema and a subsequent decrease in Brownian motion. This is manifested as a hyperintensity on DWI and a hypointensity on the corresponding ADC map. Figure~(\ref{fig:dwi_2}) is a DWI showing hyperintensity in the left MCA and PCA distributions, consistent with acute ischemic stroke. Figure~(\ref{fig:adc_stroke_1}) is an ADC map showing hypointensity in the left MCA distribution, consistent with acute ischemic stroke. Figure~(\ref{fig:adc_pca_stroke_b}) shows an ADC map with a left parieto-occipital hypointensity reflecting left posterior circulation acute ischemic stroke. In the subacute setting, the ADC may normalize and even increase due likely to ischemia-associated remodeling and loss of structural integrity. The differential for hyperintensity on DWI includes hemorrhagic stroke, traumatic brain injury, multiple sclerosis, and brain abscesses ~\cite{lima1999,bama1997,roga2005, chla2002}. DWI has long been shown in animal models to be more efficacious than $T_2$-weighted imaging for the early detection of transient cerebral ischemia~\cite{mimo1991, moco1990, crga2003}. Additionally, DWI has been shown to be highly efficacious in the early detection of acute subcortical infarctions~\cite{sich1998}. DWI in conjunction with echo-planar imaging has been shown to effectively discriminate between high grade (high cellularity) and low grade (low cellularity) gliomas~\cite{suko1999}. Of note, in spite of the relatively high specificity and sensitivity of DWI in the early detection of acute ischemic stroke, there is a small subset of stroke patients who evade DWI detection in spite of clinically evident stroke-like neurological deficits~\cite{aybu1999}. DWI tractography or diffusion tensor imaging is a form of principal component analysis in which the dominant eigendirection is used to determine the path of an axonal tract in a given voxel. This method has shown potential for further elucidation of neuronal pathways in the brain.

\subsection{Perfusion-Weighted Imaging}
\label{subsec:pwi}

Perfusion-weighted imaging (PWI) is an ordinary differential equations compartment model of blood perfusion through organs. An arterial input function (AIF) is determined as input into the particular chosen model. The AIF is an impulse, and the model is essentially described by an impulse response or Green's function. The computed output are perfusion parameters such as mean transit time, time to peak, cerebral blood flow rate, and cerebral blood volume. There are various protocols for the perfusion conditions. For instance the dynamic susceptibility contrast imaging method uses gadolinium contrast to gather local changes in $T_2*$ signal in surrounding tissue. $T_2*$ is a type of $T_2$ in which static dephasing effects are not explicitly RF-canceled, therefore dephasing from magnetic field inhomogeneities and susceptibility effects contribute to a tighter FID envelope (more rapid decay; $T_2*<T_2$) than is seen in $T_2$. Other PWI protocols include arterial spin labeling and blood oxygen level dependent labeling. Not surprisingly, PWI results vary with the choice of perfusion model and computational method with which it is implemented~\cite{yawu2002, thso2004,grdu2002,peca2002}.

\subsection{Combined Diffusion and Perfusion-Weighted Imaging}
\label{subsec:dwi_pwi}

Both PWI and DWI are efficacious in the early detection of cerebral ischemia and correlate well with various stroke quantitation scales~\cite{toye1998, bada1998, roko1998}. Furthermore, the combination of diffusion and perfusion-weighted imaging allows for assessment of the ischemic penumbra, that watershed region at the boundary of infarcted and non-infarcted tissue~\cite{neha1999, shsa2003, wuko2001, scbe1999}. It represents tissue which has suffered some degree of ischemia, but remains viable and possibly salvageable by expedient intervention with thrombolytic therapy~\cite{mato1999, paba2002, bupa2003}, hypothermia, blood pressure elevation, or other experimental methods under study. Outside the penumbra, thrombosis or autoregulation of circulation shuts down perfusion to infarcted tissue, resulting in a drastic decrease in perfusion. Diffusion is simultaneously decreased due to tissue ischemia, hence no PWI-DWI mismatch occurs. Within the penumbra however, the tissue is hypoperfused, yet since still viable, it has yet to sustain sufficient cytotoxic axonal damage to manifest a significant decrease in diffusion. The registration of both imaging modalities therefore demonstrates a diffusion-perfusion mismatch at the penumbra. In set-theoretic terms, the brain territory $\varDelta$ with compromised diffusion is a proper subset of the territory $\varPsi$ with compromised perfusion, and the relative complement of $\varDelta$ in $\varPsi$ is called the \textit{ischemic penumbra} $\varLambda$,

\begin{equation}
 \varLambda = \varDelta^c\cap\varPsi = \varPsi \setminus \varDelta 
\end{equation}

 One caveat to PWI-DWI mismatch assessment is that when done subjectively by the human eye, it can be demonstrably unreliable~\cite{cosi2003}. Hence development of quantitative metrics and learning algorithms are needed in this area. For instance, generalized linear models have been used and shown promise~\cite{wuko2001}.

\begin{figure}[ht]
\begin{center}
\scalebox{.75}
{\includegraphics{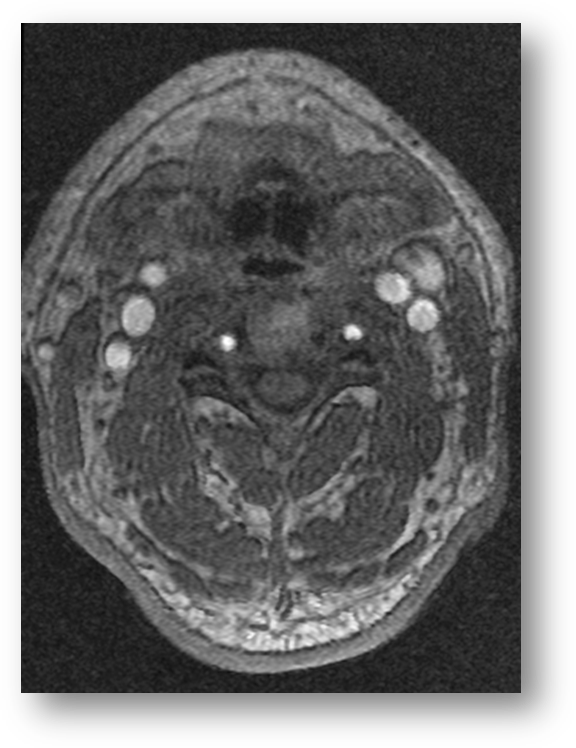}}
\end{center}
\caption{3D time-of-flight image of the carotid vasculature. Axial section of neck shown.}
\label{fig:3d_TOF}
\end{figure}

\begin{figure}[ht]
\begin{center}
\scalebox{.50}
{\includegraphics{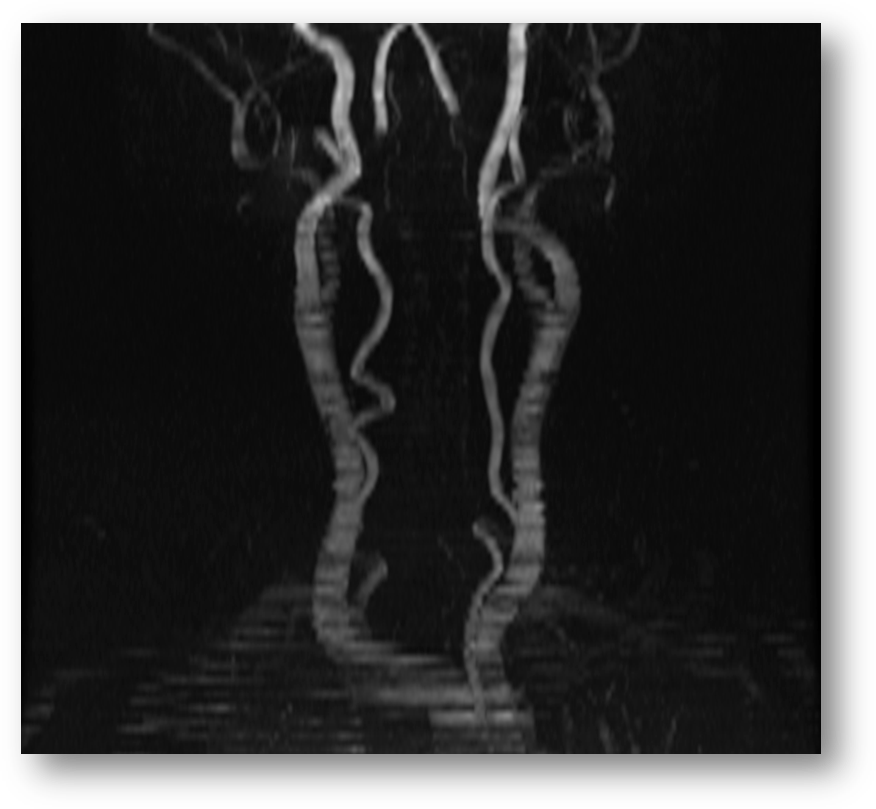}}
\end{center}
\caption{2D time-of-flight fast spoiled gradient echo sequence (FSPGR) image of the carotids.}
\label{fig:2d_tof_fspgr}
\end{figure}

\begin{figure}[ht]
\begin{center}
\scalebox{.60}
{\includegraphics{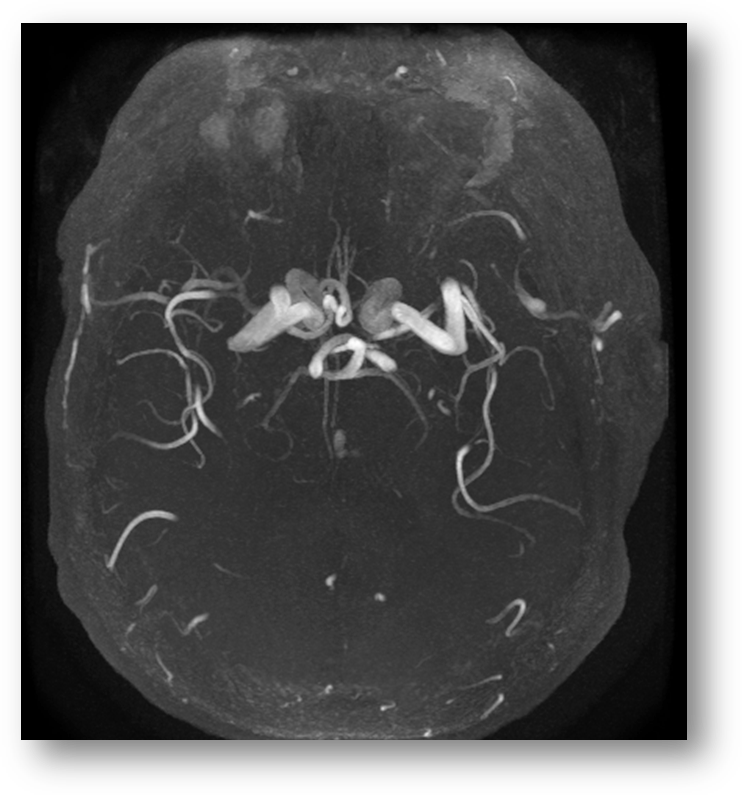}}
\end{center}
\caption{MRA showing axial view of the internal carotids and Circle-of-Willis}
\label{fig:mracowax}
\end{figure}

\begin{figure}[ht]
\begin{center}
\scalebox{.80}
{\includegraphics{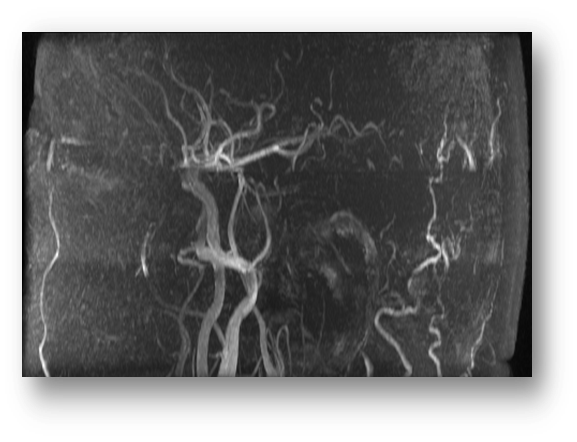}}
\end{center}
\caption{MRA lateral view showing internal carotids ascend into the Circle-of-Willis and give rise to the middle cerebral arteries. Outline of the ear is visible posterior to the internal carotids. Branches of the external carotid arteries are also seen as the facial artery anteriorly, and the arteries of the scalp posteriorly.}
\label{fig:mralat}
\end{figure}

\begin{figure}[ht]
\begin{center}
\scalebox{.50}
{\includegraphics{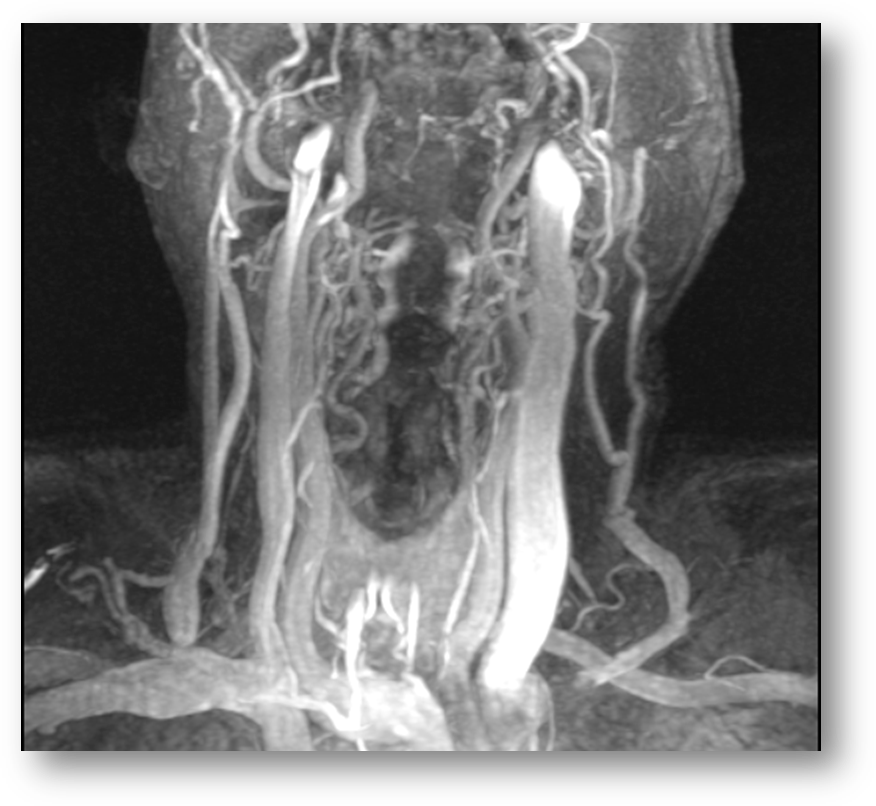}}
\end{center}
\caption{MRA of the neck showing the carotid vasculature.}
\label{fig:carotids}
\end{figure}

\subsection{Magnetic Resonance Spectroscopy}
\label{subsec:mrs}

Magnetic resonance spectroscopy (MRS) is being used in attempts to reliably measure the magnetic resonance of metabolites whose concentrations change in the setting of acute ischemic stroke. Lactate (LAC) levels are well known to increase in ischemia, and is therefore a candidate in development, while N-acetyl aspartate (NAA) levels have been shown to decrease in acute stroke, and are used in proton spectroscopy studies. Most studies confirming the ischemia-associated elevation in LAC and decrease in NAA have been done in the setting of hypoxic-ischemic encephalopathy in newborns ~\cite{cara2002, bawe2001, maju2001}. In addition to lactate, $\alpha$-Glx and glycine have also been shown to be increased in the asphyxiated neonate~\cite{mapa2002}. Phosphorus magnetic resonance spectroscopy detects changes in the resonance spectrum of energy metabolites, and has also shown prognostic significance in hypoxic-ischemic brain injury~\cite{azwy1989, pepr1985}. In their current form, MRS methods are less sensitive and practical for the assessment of acute ischemic stroke than the other magnetic resonance modalities discussed here. However, MRS is rapidly finding a place in the routine clinical assessment of hypoxic-ischemic encephalopathy in the asphyxiated neonate. Some neonatal intensive care units, for instance, now conduct MRS studies on all neonates below a certain threshold weight.

\subsection{Blood Oxygen Level-Dependent (BOLD) MRI}
\label{subsec:bold}

The blood oxygen level-dependent (BOLD) MRI is a magnetic resonance imaging method that derives contrast from the difference in magnetic properties of oxygenated versus deoxygenated blood. Hemoglobin is the molecule which carries oxygen in the blood, and is located in red blood cells. The oxygen-bound form of hemoglobin is called oxyhemoglobin, while the oxygen-free form is called deoxyhemoglobin. Deoxyhemoglobin is paramagnetic and in the presence of an applied magnetic field assumes a relatively higher magnetic dipole moment than oxyhemoglobin which is a diamagnetic molecule. Deoxygenated blood has a higher concentration of deoxyhemoglobin than oxyhemoglobin, and this difference is reflected in the MRI signal. $T_2*$-based BOLD MRI has shown promise in localizing the penumbra in acute ischemic stroke ~\cite{grka2001, gebr2006}. It is also used in functional MRI (fMRI), which is based on the principle that active brain areas have higher resource (e.g. oxygen, glucose) demands and higher waste output~\cite{hakn2003}. In this context, BOLD has been used to study the brain's behavior during sensorimotor recovery following acute ischemic stroke. Specifically, the coupling between BOLD and electrical neuroactivity has shed some light on the still poorly understood process of spontaneous motor recovery following a stroke~\cite{bise2004, caba2003, kihu2005, rope2000, crsh2006}. BOLD MRI results can be affected by baseline circulatory status, and therefore in the research setting near-infrared spectroscopy (NIRS) can be used as a control, or as an alternative modality in the clinic ~\cite{stcu2002, musa2006}. BOLD MRI has been used in assessing the brain's response to hypercapnia~\cite{vaho2005, bawo1997}. Hypercapnia can cause changes in multiple variables in the brain such as cerebral blood flow, oxygen consumption rate, cerebral blood volume, arterial oxygen concentration, and red blood cell volume fraction. However, hypercapnia-associated cerebrovascular reactivity has been strongly correlated with arterial spin labeling and other PWI surrogates, suggesting that the brain's reaction to hypercapnia is dominated by changes in cerebral blood flow~\cite{maha2008}.

\begin{figure}[ht]
\begin{center}
\scalebox{.90}
{\includegraphics{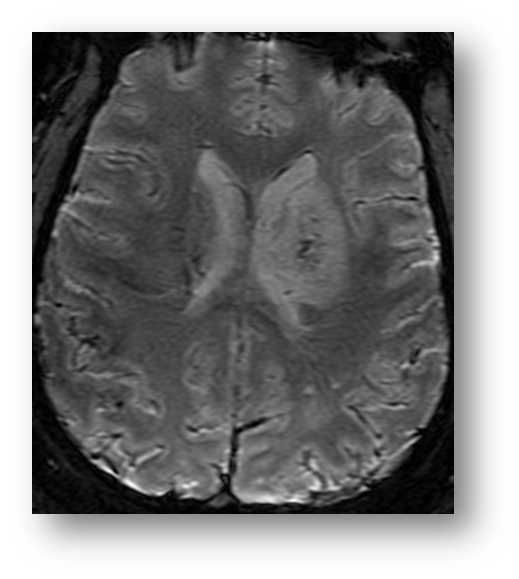}}
\end{center}
\caption{SWI showing signal in the left MCA distribution consistent with acute ischemic infarct. Note the left periventricular region of hyperintensity. Its hypointense center is consistent with the ischemic core.}
\label{fig:swi_1}
\end{figure}

\begin{figure}[ht]
\begin{center}
\scalebox{.70}
{\includegraphics{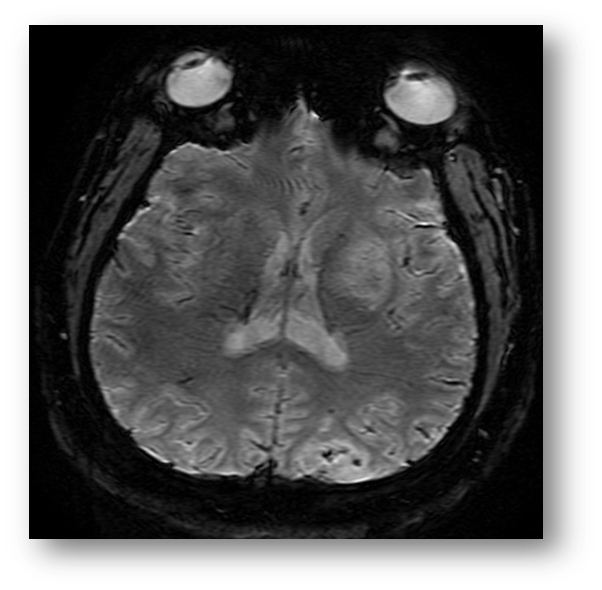}}
\end{center}
\caption{SWI showing signal in the left PCA and MCA distributions consistent with acute ischemic infarct. Note the bull's eye pattern of hyperintensity surrounding some central foci of hypointensity in the occipital lesion. }
\label{fig:swi_2}
\end{figure}
\subsection{Magnetic Resonance Angiography}
\label{subsec:mra}

MRA is a set of magnetic resonance-based techniques for imaging the circulatory system. Figures~(\ref{fig:mracowax}) and (\ref{fig:mralat}) show MRA images of the Circle-of-Willis. MRA techniques can be broadly categorized into flow-dependent and flow-independent groups. The flow-dependent methods derive contrast from the motion of blood in vasculature relative to the static state of surrounding tissue~\cite{duha1986}. Two currently well-known and used examples of flow dependent methods are: (i) Phase contrast MRA, (PC MRA)~\cite{duso1989, fina1990, prti1994} and (ii) Time-of-Flight MRA (TOF MRA). The TOF MRA images can be acquired in either two dimensional (2D TOF) or three dimensional (3D TOF) formats. Figure~(\ref{fig:2d_tof_fspgr}) shows a 2D TOF fast spoiled gradient echo sequence (FSGR) image of the carotids, and Figure~(\ref{fig:3d_TOF}) shows a an axial section of a 3D TOF image of the carotids. PC MRA exploits differences in spin phase of moving blood relative to static surrounding tissue, while TOF MRA exploits the difference in excitation pulse ($B_1$) exposure of flowing blood relative to static surrounding tissue. This difference occurs because flowing blood spends less time in the field of exposure, and as a result is less spin-saturated then the surrounding tissue. This decreased spin-saturation translates into higher intensity signals on spin-echo sequences.

Flow-independent methods exploit inherent differences in magnetic properties of blood relative to surrounding tissue. For instance, fresh blood imaging is a method that exploits the longer $T_2$ time constant in blood relative to surround. This method finds specific utility in cardiac and cardio-cerebral imaging by using fast spin echo sequences which can exploit the spin saturation differences between systole and diastole~\cite{misu2000}. Other examples include susceptibility weighted imaging (SWI) and four dimensional dynamic MRA (4D MRA). SWI derives contrast from magnetic susceptibility differences between blood and surround~\cite{heni2004}, while 4D MRA uses time-dependent bit-mask subtraction after injection of Gadolinium-DPTA or some other pharmacological contrast agent~\cite{zhbu2004}. Figures~(\ref{fig:swi_1}) and (\ref{fig:swi_2}) show SWI consistent with acute ischemic infarction in the left PCA and MCA distributions. Of note, in the sense that 4D MRA exploits the time interval between injection and initial image acquisition, it is arguably more flow-dependent than the other methods mentioned here in that category.

\section{The Hydrogen Atom}
\label{sec:H_atom}

Hydrogen is the most commonly imaged nucleus in magnetic resonance imaging, due largely to the great abundance of water in biological tissue, and to the high gyromagnetic ratio of hydrogen. Furthermore, hydrogen is the only atom whose schr\"{o}dinger eigenvalue problem has been exactly solved. Other hydrogenic atoms involve screened potentials and require iterative approximation eigenvalue problem solvers such as the Hartree-Fock scheme and its variants. In what follows we review the electronic orbital configuration of hydrogen.

\subsection{Electronic Orbital Configuration}

In this section we focus on the electron. Specifically that lone electron in the orbit of the hydrogen atom. We make this choice because its simplicity allows us review the attributes of quantum orbital angular momentum and spin, in a context which is not only real, but also highly relevant to magnetic resonance imaging, which as noted targets the hydrogen atom. The electron in $^1$H has both position and angular momentum. The Hamiltonian commutes with the squared orbital angular momentum operator, therefore angular momentum eigenstates are also energy eigenstates. Of note, the position and angular momentum observables do not commute. This is equivalent to Heisenberg's uncertainty principle. Hence definite energy or orbital angular momentum states, are represented in the position basis as probability density functions, which are synonymous with electron clouds or atomic orbitals.

Orbital angular momentum is the momentum possessed by a body by virtue of circular motion, as along an orbit. In quantum mechanics, the square of the orbital momentum is a quantized quantity, which can take on only certain discrete ``allowed'' values. This reality and discreteness of the orbital momentum eigenvalues is a manifestation of the spectral theorem of normal bounded operators; given the hermiticity and boundedness of the quantum orbital angular momentum operator over a specified finite volume such as the radius of an atom; which is itself a manifestation of the negative energy of the electron in orbit yielding a so called \textit{bound state}. The orbital momentum states can be indirectly represented by the spherical harmonic functions to be derived below. The spherical harmonics are the eigenfunctions of the squared quantum orbital angular momentum operator in spherical coordinates. Their product with the radial wave functions yield the wave function, $\psi$, of the hydrogen's electron. Where in accordance with the principal postulate of quantum mechanics, $\psi^2(r,\theta,\phi)$ is the probability of finding the electron at a given point $(r,\theta,\phi)$.

 In the time-independent case, the Hamiltonian defines the Schr\"{o}dinger equation as follows,

\begin{equation}
 H\psi = E\psi.
\label{Eqn:schrodinger}
\end{equation}

The orbital angular momentum states are the angular portion of the solution, $\psi$, of the schr\"{o}dinger equation in spherical coordinates. Substituting the expression for the Hamiltonian of a single non-relativistic particle into Equation~(\ref{Eqn:schrodinger}) yields,

\begin{equation}
 -\frac{\hbar^2}{2m}\nabla^2\psi(\bM{r}) + V(\bM{r})\psi(\bM{r}) = E\psi(\bM{r}),
\label{Eqn:schrodinger2}
\end{equation}

where $\bM{r}\in \Re^3$. In spherical coordinates the above equation becomes,

\begin{eqnarray}
 -\frac{\hbar^2r^{-2}}{2m\sin\theta} \left[\sin\theta\frac{\partial}{\partial r}\left( r^2 \frac{\partial}{\partial r} \right)+\frac{\partial}{\partial\theta} \left( \sin \theta \frac{\partial}{\partial\theta} \right) + \frac{1}{\sin\theta}\frac{\partial^2}{\partial\phi^2} \right]\psi(r) \nonumber \\ +V(r)\psi(r) = E\psi(r),
\end{eqnarray}

where $m$ denotes mass. Invoking the method of separation of variables, we assume existence of a solution of the form $\psi(r)=R(r)Y(\theta,\phi)$. We illustrate the method on the Laplace equation which can be interpreted as a homogeneous form of the schr\"{o}dinger equation, i.e. one for which $V(r)=E=0$,

\begin{eqnarray}
 \frac{Y}{r^2}\frac{\partial}{\partial r}\left( r^2 \frac{\partial}{\partial r} \right)R(r)+\frac{R}{r^2\sin\theta}\frac{\partial}{\partial\theta} \left( \sin \theta \frac{\partial}{\partial\theta} \right)Y(\theta,\phi) + \frac{R}{r^2\sin^2\theta}\frac{\partial^2Y}{\partial\phi^2} \nonumber \\ = 0.
\end{eqnarray}

Multiplying through by $r^2/RY$ yields,

\begin{eqnarray}
  \frac{1}{R}\frac{\partial}{\partial r}\left( r^2 \frac{\partial}{\partial r} \right)R(r)+\frac{1}{Y\sin\theta}\frac{\partial}{\partial\theta} \left( \sin \theta \frac{\partial}{\partial\theta} \right)Y(\theta,\phi) + \frac{1}{Y\sin^2\theta}\frac{\partial^2Y}{\partial\phi^2} \nonumber \\ = 0,
\end{eqnarray}

which can be separated as,

\begin{equation}
  \frac{1}{R}\frac{\partial}{\partial r}\left( r^2 \frac{\partial}{\partial r} \right)R(r) = \xi
\label{Eqn:radial}
\end{equation}

and 

\begin{equation}
 \frac{1}{Y\sin\theta}\frac{\partial}{\partial\theta} \left( \sin \theta \frac{\partial}{\partial\theta} \right)Y(\theta,\phi) + \frac{1}{Y\sin^2\theta}\frac{\partial^2Y}{\partial\phi^2} = -\xi.
\label{Eqn:angular}
\end{equation}

We can again assume separability in the form $Y(\theta,\phi) = \Theta(\theta)\Phi(\phi)$, and substitute into Equation~(\ref{Eqn:angular}) above to get,

\begin{equation}
  \frac{1}{\Theta\sin\theta}\frac{\partial}{\partial\theta} \left( \sin \theta \frac{\partial}{\partial\theta} \right)\Theta(\theta) + \frac{1}{\Phi\sin^2\theta}\frac{\partial^2\Phi}{\partial\phi^2} = -\xi.
\label{Eqn:SHF1}
\end{equation}

From which we extract the following two separated equations,

\begin{equation}
 \frac{1}{\Phi}\frac{\partial^2\Phi}{\partial\phi^2} = -m^2
\label{Eqn:azimuth}
\end{equation}

 and

\begin{equation}
 \xi\sin^2\theta +   \frac{\sin\theta}{\Theta}\frac{\partial}{\partial\theta} \left( \sin \theta \frac{\partial \Theta}{\partial\theta} \right) = m^2.
\label{Eqn:polar}
\end{equation}

Our homogeneous equation has therefore been separated into three ordinary differential equations: Equations~(\ref{Eqn:radial}) for the radial part, Equation~(\ref{Eqn:azimuth}) for the azimuthal part, and Equation~(\ref{Eqn:polar}) for the polar part. Solutions to the radial equation are of the form,

\begin{equation}
 R(l) = Ar^l + Br^{-(l+1)}
\end{equation}

where A and B are constant coefficients, $l$ is a non-negative integer such that $l\geq|m|$, where $m$ is an integer on the right hand side of the azimuthal and polar equations. A regularity constraint at the poles of the sphere yield a Sturm-Liouville problem which in turn mandates the form $\xi=l(l+1)$. The coefficient $A$ is often set to zero to admit only solutions which vanish at infinity. However, this choice is application-specific, and for certain applications it may be appropriate to instead set $B=0$.

Solutions to the azimuthal equation are of the form,

\begin{equation}
 Ce^{-im\phi}+De^{+im\phi},
\end{equation}

where $C$ and $D$ are constant coefficients and $e$ is the base of the natural logarithm. To obtain solutions to the polar equation, we proceed in a number of steps. First we substitute $\cos\theta\mapsto x$, and recast Equation~(\ref{Eqn:polar}) into,

\begin{equation}
 \sin^2 \theta \frac{\partial ^2\Theta}{\partial\theta^2}+ \sin\theta\cos\theta\frac{\partial\Theta}{\partial\theta} + l(l+1)\Theta\sin^2\theta -m^2\Theta  = 0,
\label{Eqn:polar2}
\end{equation}

Next we compute the derivatives of $\Theta$ under the transformation $x=\cos\theta$. Employing the chain rule yields,

\begin{equation}
 \frac{d\Theta}{d\theta} = \frac{d\Theta}{dx}\frac{dx}{d\theta} = -\sin\theta\frac{d\Theta}{dx}
\end{equation}

and

\begin{eqnarray}
  \frac{d^2\Theta}{d\theta^2}=\frac{d}{d\theta}\left(-\sin\theta\frac{d\Theta}{dx}\right)
= -\cos\theta\frac{d\Theta}{dx}-\sin\theta\frac{d}{d\theta}\frac{d\Theta}{dx}\nonumber\\
= \sin^2\theta\frac{d^2\Theta}{dx^2}-\cos\theta\frac{d\Theta}{dx}.
\end{eqnarray}

A substitution of the derivatives into the polar equation yields,

\begin{eqnarray}
 \sin^2 \theta \left(\sin^2\theta\frac{d^2\Theta}{dx^2}-\cos\theta\frac{d\Theta}{dx}\right)+ \sin\theta\cos\theta\left(-\sin\theta\frac{d\Theta}{dx}\right) \nonumber\\+ l(l+1)\Theta\sin^2\theta -m^2\Theta = 0
\label{Eqn:polar3}.
\end{eqnarray}

Dividing through by $\sin^2\theta$ and changing variables via $\cos\theta\mapsto x$ and $\Theta\mapsto y$, yields the associated ($m\neq0$) Legendre differential equation,

\begin{equation}
 (1-x^2)\frac{\partial^2 y}{\partial x^2} - 2x\frac{\partial y}{\partial x} + \left(l(l+1) - \frac{m^2}{1-x^2}\right)y = 0,
\end{equation}
 
whose solutions  are given by

\begin{equation}
 P^{m}_l(x) = \frac{(-1)^m}{2^ll!} (1-x^2)^{m/2} \frac{d^{l+m}}{dx^{l+m}}(x^2-1)^l.
\end{equation}

\begin{figure}[ht]
\begin{center}
\scalebox{.60}
{\includegraphics{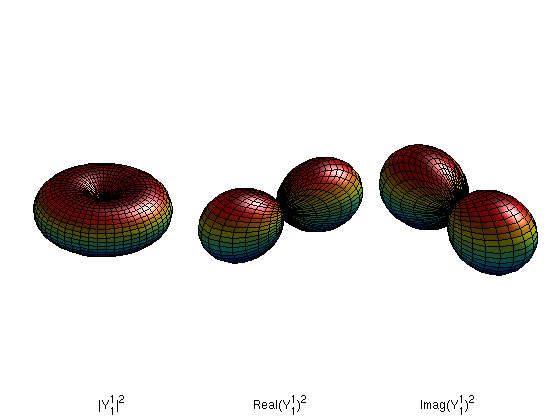}}
\end{center}
\caption{l=1,m=1}
\label{fig:SHF_1_1}
\end{figure}

\begin{figure}[ht]
\begin{center}
\scalebox{.50}
{\includegraphics{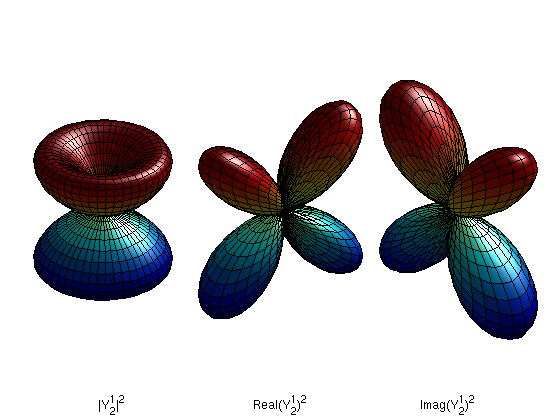}}
\end{center}
\caption{l=2,m=1}
\label{fig:SHF_2_1}
\end{figure}

\begin{figure}[ht]
\begin{center}
\scalebox{.50}
{\includegraphics{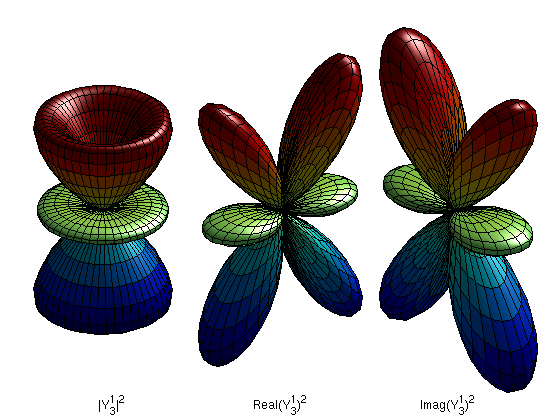}}
\end{center}
\caption{l=3,m=1}
\label{fig:SHF_3_1}
\end{figure}

\begin{figure}[ht]
\begin{center}
\scalebox{.50}
{\includegraphics{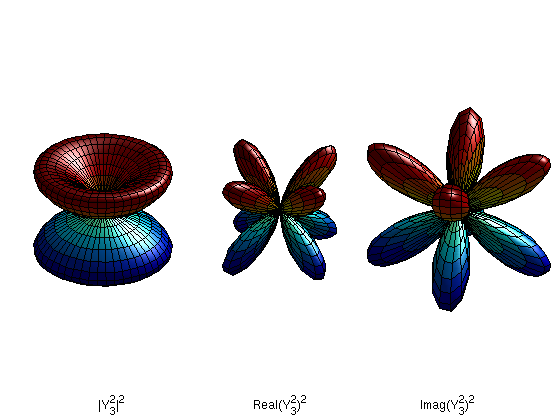}}
\end{center}
\caption{l=3,m=2}
\label{fig:SHF_3_2}
\end{figure}

\begin{figure}[ht]
\begin{center}
\scalebox{.75}
{\includegraphics{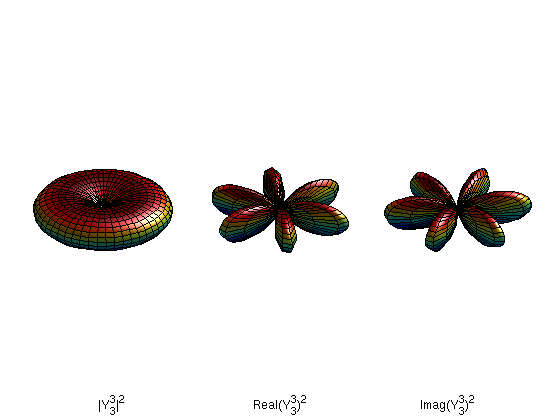}}
\end{center}
\caption{l=3,m=3}
\label{fig:SHF_3_3}
\end{figure}

\begin{figure}[ht]
\begin{center}
\scalebox{.40}
{\includegraphics{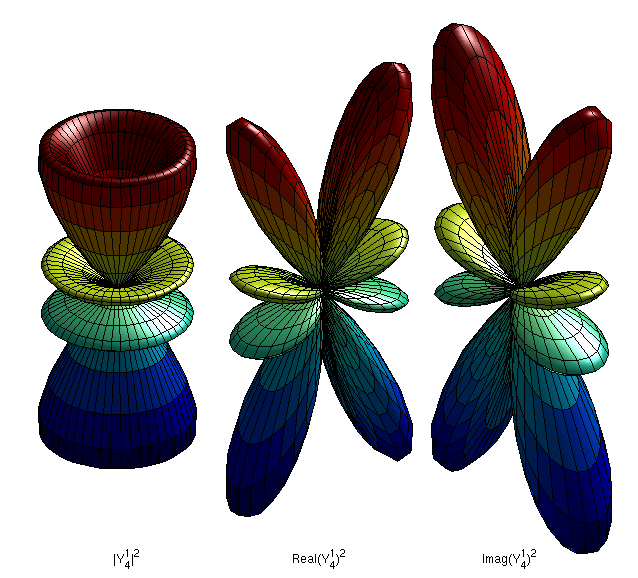}}
\end{center}
\caption{l=4,m=1}
\label{fig:SHF_4_1}
\end{figure}

\begin{figure}[ht]
\begin{center}
\scalebox{.50}
{\includegraphics{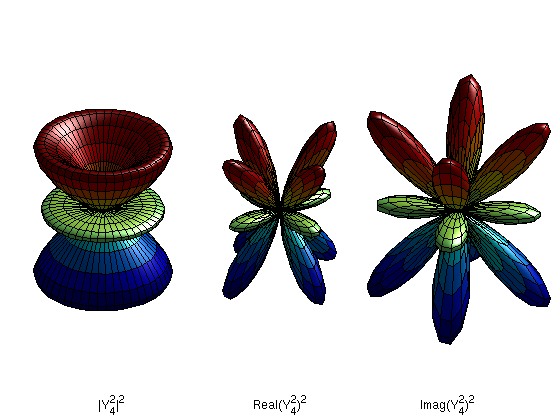}}
\end{center}
\caption{l=4,m=2}
\label{fig:SHF_4_2}
\end{figure}

\begin{figure}[ht]
\begin{center}
\scalebox{.75}
{\includegraphics{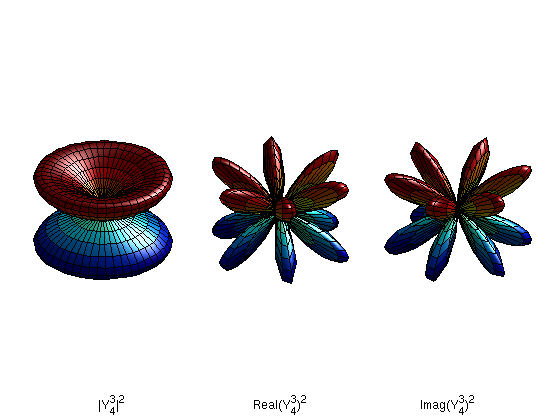}}
\end{center}
\caption{l=4,m=3}
\label{fig:SHF_4_3}
\end{figure}

\begin{figure}[ht]
\begin{center}
\scalebox{.75}
{\includegraphics{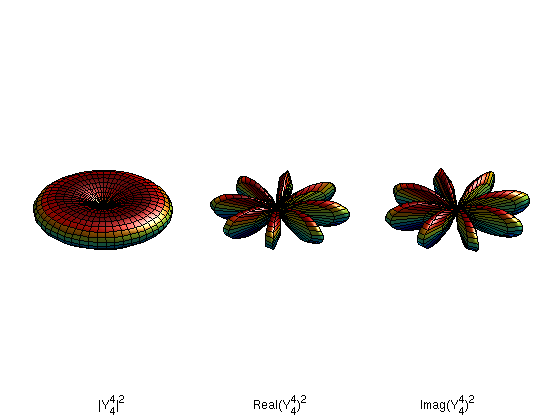}}
\end{center}
\caption{l=4,m=4}
\label{fig:SHF_4_4}
\end{figure}

\begin{figure}[ht]
\begin{center}
\scalebox{.50}
{\includegraphics{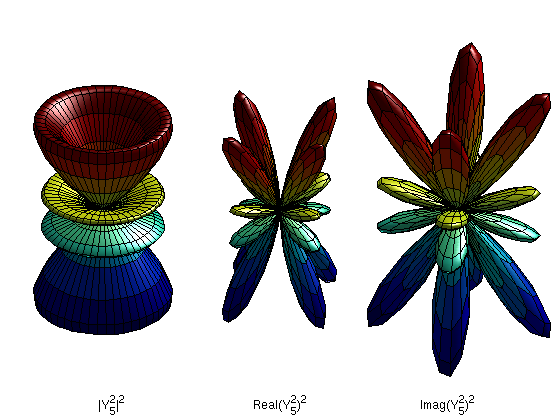}}
\end{center}
\caption{l=5,m=2}
\label{fig:SHF_5_2}
\end{figure}

The solution $Y(\phi,\theta)=\Phi(\phi)\Theta(\theta)$ to angular portion of the Laplace equation is therefore of the form,

\begin{equation}
 Y_{l,m}(\theta, \phi) = N P^m_l (\cos \theta) e^{im\phi},
\end{equation}

where N is a normalization factor given by,

\begin{equation}
 N = \sqrt{\frac{(2l+1)}{4\pi} \frac{(l-1)!}{(l+1)!}},
\end{equation}

and enabling,

\begin{equation}
 \int_{\Omega}Y_{l,m}^{*}(\Omega)Y_{l,m}(\Omega) = 1,
\end{equation}

where $\Omega=(\phi,\theta)$ is solid angle. $Y_{l,m}$ are called the spherical harmonic functions and are discussed some more in the following subsection. Figures~(\ref{fig:SHF_1_1}) to (\ref{fig:SHF_5_2}) are plots of a sampling of spherical harmonics with $1\leq l \leq 5$. 

For the electronic configuration of the hydrogen atom, the electron experiences a potential V(r) due to the proton. V(r) is the coulomb potential given by,

\begin{equation}
 V(r) = -\frac{e^2}{4\pi\epsilon_0r}
\end{equation}

Substituting this into the generic schr\"{o}dinger equation yields,

\begin{equation}
 -\frac{\hbar^2}{2m}\nabla^2\psi(r) -\frac{e^2}{4\pi\epsilon_0r}\psi(r) = E\psi(r),
\end{equation}

which in spherical coordinates is,

\begin{eqnarray}
 -\frac{\hbar^2r^{-2}}{2m\sin\theta} \left[\sin\theta\frac{\partial}{\partial r}\left( r^2 \frac{\partial}{\partial r} \right)+\frac{\partial}{\partial\theta} \left( \sin \theta \frac{\partial}{\partial\theta} \right) + \frac{1}{\sin\theta}\frac{\partial^2}{\partial\phi^2} \right]\psi(r) \nonumber \\ -\frac{e^2}{4\pi\epsilon_0r}\psi(r) = E\psi(r),
\label{Eqn:Schrodinger_EVP}
\end{eqnarray}

where $m$ is given by,

\begin{equation}
 m = \frac{m_pm_e}{m_p+m_e},
\end{equation}

and is the two-body reduced mass $m_p$ of the proton and $m_e$ of the electron.

Equation~(\ref{Eqn:Schrodinger_EVP}) above is an eigenvalue problem which after some rearranging, we can solve using the same separation of variables method illustrated above. We recast as,

\begin{eqnarray}
 \left[\frac{\partial}{\partial r}\left( r^2 \frac{\partial}{\partial r} \right)+\frac{2m}{\hbar^2}\left(\frac{re^2}{4\pi\epsilon_0} + Er^2\right)\right]R(r)Y(\theta,\psi) \nonumber \\ +\left[\frac{1}{\sin\theta} \frac{\partial}{\partial\theta} \left( \sin \theta \frac{\partial}{\partial\theta} \right) + \frac{1}{\sin^2\theta}\frac{\partial^2}{\partial\phi^2}\right]R(r)Y(\theta,\psi) \nonumber\\ = 0
\end{eqnarray}

Dividing through by $R(r)Y(\theta,\phi)$ and splitting the operator, we get,

\begin{equation}
 \frac{1}{Y(\Omega)}\left[\frac{1}{\sin\theta} \frac{\partial}{\partial\theta} \left( \sin \theta \frac{\partial}{\partial\theta} \right) + \frac{1}{\sin^2\theta}\frac{\partial^2}{\partial\phi^2}\right]Y(\Omega) = -l(l+1)
\label{Eqn:SHF2}
\end{equation}

and

\begin{equation}
 \frac{1}{R(r)} \frac{\partial}{\partial r}\left( r^2 \frac{\partial}{\partial r} \right)R(r)+\frac{2m}{\hbar^2}\left[\frac{re^2}{4\pi\epsilon_0} + Er^2\right]= l(l+1).
\label{Eqn:Laguerre}
\end{equation}

Equation~(\ref{Eqn:SHF2}) is exactly the same as Equation~(\ref{Eqn:SHF1}) which we solved above, and whose solutions are the spherical harmonic functions, $Y_{l,m}$. Equation~(\ref{Eqn:Laguerre}) above is isomorphic to a generalized Laguerre differential equation, whose solutions are related to the associated Laguerre polynomials. As mentioned in the case of Laplace's equation above, regularity conditions at the boundary prescribe the separation constant $l(l+1)$. To transform Equation~(\ref{Eqn:Laguerre}) into a generalized Laguerre equation, we recast into the following form,

\begin{equation}
  \frac{\partial}{\partial r}\left( r^2 \frac{\partial}{\partial r} \right)R(r)+\left[\frac{2mre^2}{4\pi\epsilon_0\hbar^2} + \frac{2mEr^2}{\hbar^2}- l(l+1)\right]R(r)=0.
\label{Eqn:Laguerre2}
\end{equation}

then we proceed with the following sequence of substitutions: 

\begin{eqnarray}
 R(r) \mapsto \frac{y(r)}{r}\nonumber\\
-\frac{2mE}{\hbar^2} \mapsto \frac{\beta^2}{4}\nonumber\\
r \mapsto \frac{x}{\beta}.
\label{Eqn:lagsubs1}
\end{eqnarray}

Executing the first substitution above transforms Equation~(\ref{Eqn:Laguerre2}) into,

\begin{equation}
 \frac{d^2y(r)}{dr^2} +\left[\frac{2me^2}{4\pi r\epsilon_0\hbar^2} + \frac{2mE}{\hbar^2}- \frac{l(l+1)}{r^2}\right]y(r)=0.
\label{Eqn:Laguerre3}
\end{equation}

Next, the second substitution transforms the above into,

\begin{equation}
 \frac{d^2y(r)}{dr^2} +\left[\frac{2me^2}{4\pi r\epsilon_0\hbar^2} - \frac{\beta^2}{4}- \frac{l(l+1)}{r^2}\right]y(r)=0.
\label{Eqn:Laguerre4}
\end{equation}

And finally, the third substitution transforms the above into,

\begin{equation}
 \beta^2\frac{d^2y(x)}{dx^2} +\left[\frac{2\beta me^2}{4\pi x\epsilon_0\hbar^2} - \frac{\beta^2}{4}- \frac{\beta^2l(l+1)}{x^2}\right]y(x)=0
\label{Eqn:Laguerre5}.
\end{equation}

Dividing through by $\beta^2$ yields,

\begin{equation}
\frac{d^2y(x)}{dx^2} +\left[- \frac{1}{4}+\frac{2 me^2}{4\pi \beta\epsilon_0\hbar^2x} - \frac{l(l+1)}{x^2}\right]y(x)=0
\label{Eqn:Laguerre6}.
\end{equation}

Next, we make the substitutions,

\begin{equation}
 l(l+1)\mapsto \frac{k^2-1}{4}
\label{Eqn:lagsubs2}
\end{equation}

and

\begin{equation}
\frac{2 me^2}{4\pi \beta\epsilon_0\hbar^2}\mapsto \frac{2j+k+1}{2},
\label{Eqn:lagsubs3}
\end{equation}

which transform Equation~(\ref{Eqn:Laguerre6}) into the following generalized Laguerre equation

\begin{equation}
 (y_j^k)''(x) +\left[-\frac{1}{4}+\frac{2j+k+1}{2x}-\frac{k^2-1}{4x^2}\right]y_j^k(x)=0,
\end{equation}

whose solutions are readily verified to be the following associated Laguerre functions,

\begin{equation}
 y_j^k(x) = e^{-x/2}x^{(k+1)/2}L_j^k(x),
\label{Eqn:assoc_laguerre}
\end{equation}

where $L_j^k$ are the associated Laguerre polynomials.

From Equation~(\ref{Eqn:lagsubs2}) we see that,

\begin{equation}
 k=2l+1,
\label{Eqn:k_l_relation}
\end{equation}

which we substitute into Equation~(\ref{Eqn:lagsubs3}) to yield,

\begin{equation}
 \frac{2 me^2}{4\pi \beta\epsilon_0\hbar^2}= \frac{2j+(2l+1)+1}{2}=j+l+1\equiv n.
\label{Eqn:ndef}
\end{equation}

$n$ is the principal (or radial) quantum number. The corresponding energy eigenvalues, $E_n$, are obtained by substituting $\beta^2$ into $n^2$ and solving for $E$, we get,

\begin{equation}
E_n = -\frac{\hbar^2}{2ma_0^2n^2},
\label{Eqn:Rydberg}
\end{equation}

where $a_0$ is the Bohr radius and is given by,

\begin{equation}
 a_0 = \frac{4\pi\epsilon_0\hbar^2}{m_ee^2}.
\end{equation}

The value of $E_n$ in the ground state ($n=1$) is called the Rydberg constant. Its value is shown in Table~(\ref{Tab:consts}) along with that of other physical constants pertinent to magnetic resonance imaging. The displayed values are from the 2010 Committee for Data on Science and Technology (CODATA) recommendations~\cite{mota2010}, and $u_r$ is the relative standard uncertainty.

\begin{table}[ht]
 \caption{Physical constants of magnetic resonance imaging}
\centering
\scriptsize{
\begin{tabular}{l|l|l|l|l}
\hline\hline
Quantity & Symbol & Value & Units & $u_r$  \\
\hline\hline
vacuum light speed &$c,c_0$& $299792458$ &m s$^{-1}$& exact\\
Planck constant &$h$& $6.62606957(29)\times 10^{-34}$ &J s& $4.4\times 10^{-8}$\\
Reduced Planck &$h/2\pi, ~\hbar$& $1.054571726(47)\times 10^{-34}$ &J s& $4.4\times 10^{-8}$\\
Rydberg constant& $R_\infty$& $10973731.568539(55)$ & m$^{-1}$ & $5.0\times 10^{-12}$ \\
Rydberg energy & $Ry,~hcR_\infty$& $13.605 692 53(30)$ & eV & $2.0\times 10^{-8}$ \\
Bohr radius &$a_0$& $0.52917721092(17)\times 10^{-10}$ & m& $3.2\times 10^{-10}$\\
Bohr magneton &$\mu_B$& $927.400968(20)\times 10^{-26}$ & J T$^{-1}$& $2.2\times 10^{-8}$\\
Nuclear magneton &$\mu_N$& $5.05078353(11)\times 10^{-27}$ & J T$^{-1}$& $2.2\times 10^{-8}$\\
Electron g-factor &$g_e$& $-2.00231930436153(53)$ & { }& $2.6\times 10^{-13}$\\
Proton g-factor &$g_p$& $5.585694713(46)$ & { }& $8.2\times 10^{-9}$\\
e gyromagnetic ratio & $\gamma_e$& $1.760859708(39)\times 10^{11}$& s$^{-1}$T$^{-1}$&$2.2\times 10^{-8}$\\
{ } & $\gamma_e/2\pi$& $28024.95266(62)$& MHz T$^{-1}$ & $2.2\times 10^{-8}$\\
p gyromagnetic ratio & $\gamma_p$& $2.675222005(63)\times 10^{8}$& s$^{-1}$T$^{-1}$&$2.4\times 10^{-8}$\\
{ } & $\gamma_p/2\pi$& $42.5774806(10)$& MHz T$^{-1}$ & $2.4\times 10^{-8}$\\
e magnetic moment & $\mu_e$& $-928.476430(21)\times 10^{-26}$& J T$^{-1}$&$2.2\times 10^{-8}$\\
p magnetic moment & $\mu_p$& $1.410606743(33)\times 10^{-26}$& J T$^{-1}$&$2.4\times 10^{-8}$\\
Elementary charge &$e$& $1.602176565(35)\times 10^{-19}$ & C& $2.2\times 10^{-8}$\\
Electric constant &$\epsilon_0$& $8.854187817...\times 10^{-12}$ & F m$^{-1}$& exact\\
Electron mass &$m_e$& $9.10938291(40)\times 10^{-31}$ & Kg& $4.4\times 10^{-8}$\\
Proton mass &$m_p$& $1.672621777(74)\times 10^{-27}$ & Kg& $4.4\times 10^{-8}$\\
Neutron mass &$m_n$& $1.674927351(74)\times 10^{-27}$ & Kg& $4.4\times 10^{-8}$\\
$m_p$ to $m_e$ ratio &$m_p/m_e$& $1836.15267245(75)$ & Kg& $4.1\times 10^{-10}$\\
Avogadro constant &$N_A$& $6.02214129(27)\times 10^{-23}$ & mol$^{-1}$& $4.4\times 10^{-8}$\\
Molar gas const &$R$& $8.3144621(75)$ & J mol$^{-1}$ K$^{-1}$& $4.4\times 10^{-8}$\\
Boltzmann constant &$R/N_A,~k$& $1.3806488(13)\times 10^{-23}$ & J K$^{-1}$& $9.1\times 10^{-7}$\\
Electron volt &(J/C)e,  eV &$1.602176565(35)\times 10^{-19}$ & J& $2.2\times 10^{-8}$\\
\hline
\end{tabular}
}
 \label{Tab:consts}
\end{table}

To transform the radial equation solution, Equation~(\ref{Eqn:assoc_laguerre}), into a form in terms of radial distance $r$, and principal and azimuthal quantum numbers $n$ and $l$, we note the following relations:

From Equation~(\ref{Eqn:k_l_relation}) we get,
\begin{equation}
 \frac{k+1}{2} = \frac{(2l+1)+1}{2} = l+1;
\end{equation}
 
from Equation~(\ref{Eqn:ndef}) we get,
\begin{equation}
 j+l+1\Rightarrow j = n-l-1;
\end{equation}

and from Equations~(\ref{Eqn:lagsubs1}) and (\ref{Eqn:Rydberg}) we get,
\begin{equation}
 \frac{\beta^2}{4}=-\frac{2mE_n}{\hbar^2}=-\frac{2m}{\hbar^2}\left(-\frac{\hbar^2}{2ma_0^2n^2}\right) = \frac{1}{(a_0n)^2}\Rightarrow \beta=\frac{2}{a_0n},
\end{equation}

Therefore,

\begin{equation}
 x=\beta r = \frac{2r}{a_0n}.
\end{equation}

Substituting the above derived expressions of $x$, $j$, and $k$ into the associated Laguerre function, Equation~(\ref{Eqn:assoc_laguerre}), we get,

\begin{equation}
 y_{n-l-1}^{2l+1}(r)=e^{-r/a_0n}\left(\frac{2r}{a_0n}\right)^{l+1}L^{2l+1}_{n-l-1}\left(\frac{2r}{a_0n}\right).
\end{equation}

Next we make the substitution $R=y(r)/r$, which yields the radial solutions of the hydrogen atom,

\begin{equation}
R(r)=C e^{-r/a_0n}\left(\frac{2r}{a_0n}\right)^{l}L^{2l+1}_{n-l-1}\left(\frac{2r}{a_0n}\right),
\end{equation}

where the constant $C=\frac{2}{a_0n} $. Following multiplication by the spherical harmonics and normalization, we obtain the exact solution of the time-independent Schr\"{o}dinger equation of the hydrogen atom,

\begin{eqnarray}
\langle r,\theta,\phi|n,l,m\rangle=\Psi(r,\theta,\phi) = \nonumber\\ \sqrt{\left(\frac{2}{na_0}\right)^3\frac{(n-l-1)!}{2n[(n+1)!]^3}}e^{-r/na_0}\left(\frac{2r}{na_0}\right)^lL^{2l+1}_{n-l-1}\left(\frac{2r}{na_0}\right)Y_{l,m}(\theta,\phi),
\end{eqnarray}

where the principal, azimuthal, and magnetic quantum numbers ($n$, $l$, $m$) take the following values,

\begin{eqnarray*}
 n = 1,2,3...\\
l=0,1,2,...n-1\\
m=-l,-l+1,...0,...l-1,l
\end{eqnarray*}

\subsection{Spherical Harmonics and Radial Wave Functions}
\label{subsec:SHF_RWF}

In this section, we review the spherical harmonics and the radial wave functions. As noted above, the spherical harmonics, $Y_l^{m_l}$, are characterized by an azimuthal and a magnetic quantum number, $l$ and $m_l$ respectively. Table~(\ref{Tab:SHF}) shows the spherical harmonics with $l$ = 0,1,2, and 3. 
Formalism of state representation by the spherical harmonics is as follows,

\begin{equation}
 |l,m\rangle = Y_{l,m}
\end{equation}

\begin{equation}
 \langle \theta, \phi | l, m\rangle = Y_{l,m}(\theta, \phi)
\end{equation}

where, $0\leq \theta \leq \pi$ is the polar angle and $0 \leq \phi \leq 2\pi$ is the azimuthal angle. The Spherical harmonic representing rotation quantum numbers  $| l, m\rangle$ at position state $|\theta, \phi\rangle$, is given by

\begin{equation}
 Y_{l,m}(\theta, \phi) = \sqrt{\frac{(2l+1)}{4\pi} \frac{(l-1)!}{(l+1)!}} P^m_l (\cos \theta) e^{im\phi}.
\end{equation}

\begin{table}[ht]
 \caption{Spherical Harmonics for $l=0,1,2,\mbox{and}~3$.}
\centering
\small{
\begin{tabular}{c|c|l}
\hline\hline
$l$ & $m_l$ & $Y_l^{m_l}$ \\
\hline\hline
0 & 0 & $Y_0^0(\theta,\phi)=\frac{1}{2}\sqrt{\frac{1}{\pi}}$ \\
1 & -1 &$Y_{1}^{-1}(\theta,\phi)=\frac{1}{2}\sqrt{\frac{3}{2\pi}}\sin\theta e^{-i\phi}$ \\
1 & 0 &$Y_{1}^{0}(\theta,\phi)=\frac{1}{2}\sqrt{\frac{3}{\pi}}\cos\theta$ \\
1 & 1 & $Y_{1}^{1}(\theta,\phi)=-\frac{1}{2}\sqrt{\frac{3}{2\pi}}\sin\theta e^{i\phi}$ \\
2 & -2 & $Y_{2}^{-2}(\theta,\phi)=\frac{1}{4}\sqrt{\frac{15}{2\pi}}\sin^2\theta e^{-2i\phi}$ \\
2 & -1 & $Y_{2}^{-1}(\theta,\phi)=\frac{1}{2}\sqrt{\frac{15}{2\pi}}\sin\theta \cos\theta e^{-i\phi}$ \\
2 & 0 & $Y_{2}^{0}(\theta,\phi)=\frac{1}{4}\sqrt{\frac{5}{\pi}}(3\cos^2\theta -1)$ \\
2 & 1 &$Y_{2}^{1}(\theta,\phi)=-\frac{1}{2}\sqrt{\frac{15}{2\pi}}\sin\theta\cos\theta e^{i\phi}$\\
2 & 2 & $Y_{2}^{2}(\theta,\phi)=\frac{1}{4}\sqrt{\frac{15}{2\pi}}\sin^2\theta e^{2i\phi}$ \\
3 & -3 &$Y_{3}^{-3}(\theta,\phi)=\frac{1}{8}\sqrt{\frac{35}{\pi}}\sin^3\theta e^{-3i\phi}$\\
3 & -2 &$Y_{3}^{-2}(\theta,\phi)=\frac{1}{4}\sqrt{\frac{105}{2\pi}}\sin^2\theta\cos\theta e^{-2i\phi}$\\
3 & -1 &$Y_{3}^{-1}(\theta,\phi)=\frac{1}{8}\sqrt{\frac{21}{\pi}}(5\cos^2\theta -1)\sin\theta e^{-i\phi}$\\
3 &  0 &$Y_{3}^{0}(\theta,\phi)=\frac{1}{4}\sqrt{\frac{7}{\pi}}(5\cos^3\theta -3\cos\theta)$\\
3 & 1 &$Y_{3}^{1}(\theta,\phi)=-\frac{1}{8}\sqrt{\frac{21}{\pi}}(5\cos^2\theta-1)\sin\theta e^{i\phi}$\\
3 & 2 &$Y_{3}^{2}(\theta,\phi)=\frac{1}{4}\sqrt{\frac{105}{2\pi}}\sin^2\theta\cos\theta e^{2i\phi}$\\
3 & 3 &$Y_{3}^{3}(\theta,\phi)=-\frac{1}{8}\sqrt{\frac{35}{\pi}}\sin^3\theta e^{3i\phi}$\\
\hline
\end{tabular}
}
 \label{Tab:SHF}
\end{table}

\begin{table}[h]
 \caption{Radial Wave Functions for $n=0,1,2,\mbox{and}~3$.}
\centering
\small{
\begin{tabular}{c|c|l}
\hline\hline
$n$ & $l$ & $R_n^l$ \\
\hline\hline
1 & 0 & $R_1^0(r)=2\left(\frac{1}{a_0}\right)^{3/2}e^{-r/a_0}$ \\
2 & 0 & $R_2^0(r)=\left(\frac{1}{2a_0}\right)^{3/2}(2-r/a_0)e^{-r/2a_0}$ \\
2 & 1 & $R_2^1(r)=\left(\frac{1}{2a_0}\right)^{3/2}\frac{r}{a_0\sqrt{3}} e^{-r/2a_0}$ \\
3 & 0 & $R_3^0(r)=2\left(\frac{1}{3a_0}\right)^{3/2}\left(1-\frac{2r}{3a_0}+\frac{2}{27}(r/a_0)^2\right)e^{-r/3a_0}$ \\
3 & 1 & $R_3^1(r)=\left(\frac{1}{3a_0}\right)^{3/2}\frac{4\sqrt{2}}{3}\left(1-\frac{r}{6a_0}\right)e^{-r/3a_0}$ \\
3 & 2 & $R_3^2(r)=\left(\frac{1}{3a_0}\right)^{3/2}\frac{2\sqrt{2}}{27\sqrt{5}}\left(\frac{r}{a_0}\right)^2 e^{-r/3a_0}$ \\
\hline
\end{tabular}
}
 \label{Tab:RadWavs}
\end{table}

\subsubsection{Operator Representation}

The spherical harmonics are a complete set of orthonormal functions over the unit sphere. In particular, they are eigenfunctions of the square of the orbital angular momentum operator, $L^2$. And are thereby representations of the allowed discrete states of angular momentum. $L^2$ can be arrived at by representing Laplace's equation in spherical coordinates and considering only its angular portion. Classically, angular momentum is:

\begin{equation}
 \bM{L=r\times p},
\end{equation}

where $\bM{r}$ is the position vector and $\bM{p}$ is the momentum vector.

By analogy, the quantum orbital angular momentum operator is given by,

\begin{equation}
 \bM{L} = -i\hbar ~(\bM{r\times \nabla}),
\end{equation}

where $\nabla$ is the gradient operator, and we have used the momentum operator of quantum mechanics,

\begin{equation}
 \bM{p} = -i\hbar \nabla.
\end{equation}

In spherical coordinates, $L^2$ is then represented by,

\begin{equation}
 L^2 = -\hbar^2 \left(\frac{1}{\sin \theta}\frac{\partial}{\partial\theta} \left( \sin \theta \frac{\partial}{\partial\theta} \right) + \frac{1}{\sin^2\theta} \frac{\partial^2}{\partial\phi^2} \right),
\end{equation}

and similarly, $L_x$, $L_y$, and $L_z$ are represented as follows,

\begin{equation}
 L_x = i\hbar \left(\sin \phi \frac{\partial}{\partial \theta} + \cot \theta \cos \phi \frac{\partial}{\partial \phi} \right),
\end{equation}

\begin{equation}
 L_y = i\hbar \left( - \cos \phi \frac{\partial}{\partial \theta} + \cot \theta \sin \phi \frac{\partial}{\partial \phi} \right),
\end{equation}

\begin{equation}
 L_z = -i\hbar \frac{\partial}{\partial \phi}.
\end{equation}

Given the expressions above for $Y_{l,m}(\theta,\phi)$, $L^2$, and $L_z$, it is readily shown that,

\begin{equation}
 L^2|l,m\rangle = \hbar^2 l(l+1)|l,m\rangle,
\end{equation}

and

\begin{equation}
 L_z|l,m\rangle = \hbar m |l,m\rangle.
\end{equation}

The square of the radial wave functions is the probability the electron is located a distance $r$ from the nucleus. Table~(\ref{Tab:RadWavs}) shows the radial wave functions for $n$ = 1,2, and 3.

\subsubsection{Commutation Relations and Ladder Operators}

\begin{equation}
 [L_i,L_j] = i\hbar \epsilon_{ijk}L_k, 
\end{equation}

where $\epsilon_{ijk}$ is the Levi-Civita symbol given by,

\begin{equation}
 \epsilon_{ijk} = \begin{cases}
\displaystyle +1 &~\mbox{if (i,j,k) is an even permutation of (1,2,3)},\\
\displaystyle -1 &~\mbox{if (i,j,k) is an odd permutation of (1,2,3}),\\
\displaystyle 0 &~\mbox{if any index is repeated},\\
\end{cases} 
\end{equation}

and i,j, k can have values of x, y, or z.

\begin{equation}
 [L^2,L_j] = 0 \qquad \mbox{for} ~j=x,y,z
\end{equation}

The ladder operators act to increase or decrease the $m$ quantum number of a state and are given by,

\begin{equation}
L_{\pm} = L_x \pm iL_y 
\end{equation}

It is readily verified that,

\begin{equation}
 L_{\pm}|l,m\rangle = \hbar \sqrt{(l\mp m)(l+1\pm m)}|l,m\pm1\rangle,
\end{equation}

\begin{equation}
[L_z,L_{\pm}] = \pm \hbar L_{\pm},
\end{equation}

and

\begin{equation}
 [L_{+},L_-] = 2\hbar L_z.
\end{equation}

\section{Intrinsic Spin}
\label{sec:spin}

Spin is an intrinsic quantum property of elementary particles such as the electron and the quark. Composite particles such as the neutron, proton, and even atoms and molecules also possess spin by virtue of their composition from their elementary particle constituents. The term spin is itself a misnomer, as a physically spinning object about an axis is not sufficient to account for the observed magnetic moments. The mathematics of spin was worked out by Wolfgang Pauli, who either astutely or serendipitously neglected to name it. This was wise, as later insight elucidated spin as an intrinsic quantum mechanical property with no classical correlate.

Spin follows essentially the same mathematics as outlined above for orbital angular momentum. This is by virtue of the isomorphism of the respective Lie Algebras of SO(3) and SU(2) groups. And is elaborated further in section~(\ref{sec:groups}) below. One notable difference is that half-integer eigenvalues are allowed for spin, while orbital angular momentum admits only integer values. The spin algebra is encapsulated in the commutation relations as follows,

\begin{equation}
  [S_i,S_j] = i\hbar \epsilon_{ijk}S_k, 
\end{equation}
 
\begin{equation}
 [S^2,S_j] = 0 \qquad \mbox{for} ~j=x,y,z,
\end{equation}

\begin{equation}
S_{\pm} = S_x \pm iS_y ,
\end{equation}

\begin{equation}
 S_{\pm}|s,m\rangle = \hbar \sqrt{(s\mp m)(s+1\pm m)}|s,m\pm1\rangle,
\label{Eqn:spinladder}
\end{equation}

\begin{equation}
[S_z,S_{\pm}] = \pm \hbar S_{\pm},
\end{equation}

and

\begin{equation}
 [S_{+},S_-] = 2\hbar S_z.
\end{equation}

\section{Addition and the Clebsch-Gordan Coefficients}
\label{sec:adding}

The addition of spin and or of orbital angular momentum refers to the process of determining the net spin and or orbital angular momentum of a system of particles. For example, the spin of a hydrogen atom in its ground state, i.e. the $l=0$ state with zero orbital angular momentum, constitutes the composite spin of the electron and the proton. Similarly, the nuclear spin, $I$, is the composition of the spins of protons and neutrons in the nucleus of an atom. 

Mathematically, the process is described by a change of basis. A change from the product space basis $\langle j_1,m_1\lvert\langle j_2,m_2\lvert$ to the ``net sum'' space basis $\langle j,m\lvert$, where $m=m_1+m_2$ and $|j_1-j_2|\leq j\leq j_1+j_2$. The representation of the sum space eigenstates in terms of the product space eigenstates are referred to as the \textit{Clebsch-Gordan coefficients}.

\begin{equation}
 |j,m\rangle = \sum_{m_1=-j_1}^{j_1}\sum_{m_2=-j_2}^{j_2} C(m_1,m_2,m)|j_1,m_1\rangle |j_2,m_2\rangle,
\end{equation}

where $C(m_1,m_2,m)$ are the Clebsch-Gordan coefficients and are given by,

\begin{equation}
 C(m_1,m_2,m) = \langle j,m| j_1 j_2, m_1 m_2\rangle,
\end{equation}
and we have notationally represented the uncoupled product space basis by,

\begin{equation}
 |j_1,m_1\rangle |j_2,m_2\rangle := | j_1 j_2, m_1 m_2\rangle.
\end{equation}

\subsection{Addition Algorithm}

Here we review a procedural description of the addition of angular momenta. Consider a hydrogen atom $^1$H in the ground state. The spin contributors are the lone electron in the l=0 shell, and a proton which is the sole constituent of the nucleus. The possible configurations of spin are:

\begin{equation}
 |\uparrow\uparrow\rangle, ~|\uparrow\downarrow\rangle, ~|\downarrow\uparrow\rangle, ~\mbox{and}~|\uparrow\uparrow\rangle 
\end{equation}

where $\uparrow$ indicates spin $+1/2$ (spin up) and $\downarrow$ indicates spin $-1/2$ (spin down). And where each of the above states are prepared by measuring $S^2_1$, $S^2_2$, $S_{z,1}$ and $S_{z,2}$ for each state. We are interested in the combined state of the two particle system, and this requires the quantum mechanical addition of angular momenta, or in this example, spin. The net result is a state $|s,m\rangle$ whose spin can be determined by measuring $S^2$ and whose $z$-component of spin can be determined by measuring $S_z$, where $S = S_1+S_2$ and $S_z = S_{z,1}+S_{z,2}$. It follows that $m=m_1+m_2$, and therefore the possible values of $m$ are 1, 0, 0, -1. This suggests a triplet state (s=1) and a singlet state (s=0). The triplet state corresponds to m=1,0,-1, and is given by,

\begin{equation}
\begin{array}{ll}
|1,1\rangle~~\mapsto &|\uparrow\uparrow\rangle \\
|1,0\rangle~~\mapsto &\frac{1}{\sqrt{2}}\left(|\uparrow\downarrow\rangle + |\downarrow\uparrow\rangle\right)\\
|1,-1\rangle\mapsto &|\downarrow\downarrow\rangle
\end{array}
\end{equation}

and the singlet state corresponds to,

\begin{equation}
 |0,0\rangle~~\mapsto \frac{1}{\sqrt{2}}(|\uparrow\downarrow\rangle - |\downarrow\uparrow\rangle),
\end{equation}

where upon designation of the top assignment $|1,1\rangle$, each of the other states are obtained by application of the ladder operators defined in Equation~(\ref{Eqn:spinladder}). For example,

\begin{equation}
 S_-|1,1\rangle = \hbar\sqrt{2}|1,0\rangle
\end{equation}

To change to a representation in the uncoupled basis, we write,

\begin{equation}
 |1,0\rangle=\frac{1}{\hbar\sqrt{2}}S_-|1,1\rangle = \frac{1}{\hbar\sqrt{2}}(S_{1,-}+S_{2,-})|\uparrow\uparrow\rangle,
\end{equation}

and therefore,

\begin{equation}
 |1,0\rangle = \frac{1}{\hbar\sqrt{2}}(S_{1,-}|\uparrow\uparrow\rangle + S_{2,-}|\uparrow\uparrow\rangle),
\end{equation}

which yields,

\begin{equation}
 |1,0\rangle = \frac{1}{\hbar\sqrt{2}}(|\downarrow\uparrow\rangle + |\uparrow\downarrow\rangle),
\end{equation}

Finally to obtain the singlet state, we simply orthogonalize the above, yielding,

\begin{equation}
|0,0\rangle =  \frac{1}{\hbar\sqrt{2}}(|\downarrow\uparrow\rangle - |\uparrow\downarrow\rangle)
\end{equation}

The above procedure is easily carried out by a computer for any combination of particle spins. For many particle systems, for instance in the computation of nuclear spins or non-ground state $^1$H configurations, the associativity property is used and the above description applies. In the above example, the factors $\frac{1}{\hbar\sqrt{2}}$ are the Clebsch-Gordan coefficients. There are several openly available implementations of Clebsch-Gordan coefficient calculators, in addition to tabulations in handbooks of physics formulae~\cite{wo2000, abst1964}.

\section{Group Theory: SO(3), SU(2), and SU(3)}
\label{sec:groups}

Orbital angular momentum and spin are the source of magnetic resonance. Their abstract mathematical description is group theoretic, and extends naturally into other fundamental physics such as the strong interaction of quarks in nucleons and nucleons in nuclei. The Frenchman Henri Poincar\'{e} commented that ``mathematicians do not study objects, but the relationships between objects''. Relationships between symmetry groups have indeed been the essential device for probing the unseen in the realm of high energy particle physics.

For spin 1/2 particles such as the electron and the quark, the spin observable can be represented by the SU(2) group. SU(N) is the group of $N\times N$ unitary matrices with unit determinant, in which the group operation is matrix-matrix multiplication. Similarly SO(N) is the group of $N\times N$ orthogonal matrices of unit determinant, with matrix multiplication as group operator.  Analogous to the SU(2) representation of spin, quantum orbital angular momentum can be represented by the SO(3) group. Both groups share a Lie Algebra, encoded in the commutation relations shown above, because they are isomorphic to each other. In particular, SU(2) is a double cover of SO(3). The SU(3) group models the interactors of the strong force in Quantum Chromodynamics (QCD). 

\subsection{SO(3)}

A counterclockwise rotation by an angle $\phi$ about the x, y, or z-axes, can respectively be represented by

\begin{equation}
\begin{array}{ccc}
 
R_x(\phi) = \left( \begin{array}{ccc}
1 & 0 & 0\\
0 &\cos\phi & -\sin\phi\\
0 &\sin\phi & \cos\phi\end{array} \right),

\\
\\

R_y(\phi) = \left( \begin{array}{ccc}
\cos\phi & 0& \sin\phi\\
0 & 1 & 0\\
-\sin\phi & 0 &\cos\phi\end{array} \right),

\\
\\

R_z(\phi) = \left( \begin{array}{ccc}
\cos\phi & -\sin\phi&0\\
\sin\phi & \cos\phi&0\\
0&0&1\end{array} \right).

\end{array}
\label{rotation}
\end{equation}

The corresponding infinitesimal generators, $G_x$, $G_y$, $G_z$ respectively are:

\begin{equation}
\begin{array}{rrr}
 
\left( \begin{array}{ccc}
0 & 0 & 0\\
0 &0 & -1\\
0 &1 & 0\end{array} \right),

&

\left( \begin{array}{ccc}
0 & 0& 1\\
0 & 0 & 0\\
-1 & 0 &0\end{array} \right),

&

 \left( \begin{array}{ccc}
0 & -1&0\\
1 & 0&0\\
0&0&0\end{array} \right).

\end{array}
\label{so3_generators}
\end{equation}

The corresponding Lie Algebra can then be readily shown to be:

\begin{equation}
 [G_i,G_j] = i\hbar \epsilon_{ijk} G_k
\end{equation}

\subsection{SU(2)}

\begin{equation}
 SU(2) =  \left\{ \left( \begin{array}{cc}
a & -b^* \\
b &a^*\end{array} \right): a, b \in \bM{C}, a^2+b^2=1 \right\}
\end{equation}

It follows that the generators of SU(2) are the Pauli Matrices given by,

\begin{equation}
\begin{array}{rrr}
 
\sigma_1=\left( \begin{array}{cc}
0 & 1 \\
1 &0\end{array} \right),

&

\sigma_2=\left( \begin{array}{cc}
0 & -i \\
i &0\end{array} \right),

&

\sigma_3=\left( \begin{array}{cc}
1 & 0 \\
0 &-1\end{array} \right),

\end{array}
\label{su2_generators}
\end{equation}

The SU(2) Lie Algebra is then prescribed by the following commutation and anti-commutation relations,

\begin{equation}
 [\sigma_a,\sigma_b] = 2i\epsilon_{abc}\sigma_c ,
\end{equation}

and

\begin{equation}
 \{\sigma_a,\sigma_b\} = 2\delta_{ab}\cdot I + i\epsilon_{abc}\sigma_c, 
\end{equation}

which combine to give:

\begin{equation}
 \sigma_a \sigma_b = \delta_{ab}\cdot I + i\epsilon_{abc}\sigma_c,
\end{equation}

where $\delta_{ab}$ is the Kronecker delta and $\{x,y\}:=xy+yx$ is the anti-commutator.

\subsubsection{SU(2) is Isomorphic to the 3-Sphere}

 Given $U \in ~SU(2)$, such that,

\begin{equation}\nonumber
 U=\left( \begin{array}{cc}
a & -b^* \\
b &a^*\end{array} \right),
\end{equation}

We define an assignment, $\zeta$, such that

\begin{equation}
\zeta: \begin{cases}
 a \mapsto x_0 + ix_3 \\
b \mapsto -x_2+ix_1
\end{cases},
\end{equation}

where $x_0, x_1, x_2, x_3 \in \Re$. It follows that,

\begin{equation}
 U = x_0I + ix_p\sigma_p,
\end{equation}

where $I$ is the $2\times2$ identity matrix, $\sigma_p$ are the Pauli matrices, and we have used the Einstein summation notation over $p$.

\begin{equation}
 det(U) = 1\Rightarrow a^2+b^2=1\Rightarrow x_0^2+x_px_p = 1 \Rightarrow \bM{x}\in S^3
\end{equation}

$\zeta$ is a bijective homomorphism between $SU(2)$ and $S^3$. And this shows $SU(2)$ is isomorphic to $S^3$.

\subsection{SU(3)}

The generators of the SU(3) group are given by the Gell-Mann matrices:

\begin{equation}
\begin{array}{cc}
\begin{array}{rr}
 
\lambda_1=\left( \begin{array}{ccc}
0 & 1 & 0\\
1 &0 & 0\\
0 &0 & 0\end{array} \right),

&

\lambda_2=\left( \begin{array}{ccc}
0 &-i & 0\\
i &0 & 0\\
0 &0 & 0\end{array} \right),

\end{array}

\\
\\
\begin{array}{rr}
 
\lambda_3=\left( \begin{array}{ccc}
1 & 0 & 0\\
0 &-1 & 0\\
0 &0 & 0\end{array} \right),

&

\lambda_4=\left( \begin{array}{ccc}
0 &0 & 1\\
0 &0 & 0\\
1 &0 & 0\end{array} \right),

\end{array}

\\
\\
\begin{array}{rr}
 
\lambda_5=\left( \begin{array}{ccc}
0 & 0 & -i\\
0 &0 & 0\\
i &0 & 0\end{array} \right),

&

\lambda_6=\left( \begin{array}{ccc}
0 &0 & 0\\
0 &0 & 1\\
0 &1 & 0\end{array} \right),

\end{array}

\\
\\
\begin{array}{rr}
 
\lambda_7=\left( \begin{array}{ccc}
0 & 0 & 0\\
0 &0 & -i\\
0 &i & 0\end{array} \right),

&

\lambda_8=\frac{1}{\sqrt{3}}\left( \begin{array}{ccc}
1 &0 & 0\\
0 &1 & 0\\
0 &0 & -2\end{array} \right),

\end{array}
\end{array}
\end{equation}

Then defining $T_j = \frac{\lambda_j}{2}$, the commutation relations are,

\begin{equation}
 [T_a,T_b] = i\sum_{c=1}^8f_{abc}T_c ,
\end{equation}

and

\begin{equation}
 \{T_a,T_b\} = \frac{4}{3}\delta_{ab} + 2\sum_{c=1}^8d_{abc}T_c, 
\end{equation}

where $f_{abc}$ and $d_{abc}$ are the antisymmetric and symmetric structure constants of SU(3) respectively.

\section{Summary}
\label{sec:summary}

Magnetic resonance imaging is an application of quantum mechanics which has revolutionized the practice of medicine. Electrons, protons, and neutrons are magnets by virtue of their spin and orbital angular momentum. Tissue is made of ensembles of such subatomic particles, and can be imaged by sensing their responses to applied magnetic fields. As discussed in this paper, a wide array of MRI modalities exist, each of which derives contrast by specific perturbations of the magnetization vector. The potential of MRI in the diagnosis and even the treatment of various diseases is far from fully realized. A quantum level understanding of magnetic resonance technology is essential for further innovation in this field.

\newpage
\section*{Acknowledgement}

The author thanks his friends Henok T. Mebrahtu and Peter Q. Blair for reading this manuscript and providing helpful feedback. And he thanks them for many enjoyable physics conversations over coffee. He thanks Daniel Ennis for making available his MATLAB code for plotting spherical harmonics. He thanks his lovely wife Lisa M. Odaibo for a helpful conversation on the clinical use of magnetic resonance spectroscopy in hypoxic-ischemic injury assessment in the neonate. And he especially thanks her for her love and support. The author thanks his entire family, and celebrates his father, Dr. Stephen K. Odaibo, on his birthday today. He thanks him for being a kind wonderful father and example. He thanks those who contributed and are contributing to his education, including: Dr. Robert A. Copeland Jr., Dr. Leslie Jones, Dr. Janine Smith-Marshall, Dr. Bilal Khan, Dr. David Katz, Dr. Earl Kidwell, Dr. William Deegan, Dr. Brian Brooks, Dr. Natalie Afshari,  Dr. Isaac O. Karikari, Dr. Xiaobai Sun, Dr. Mark Dewhirst, Dr. Carlo Tomasi, Dr. John Kirkpatrick, Dr. Nicole Larrier, Dr. Brenda Armstrong, Dr. Phil Goodman, Dr. Joanne Wilson and Dr. Ken Wilson, Dr. Srinivasan Mukunduan Jr., Dr. David Simel, Dr. Danny O. Jacobs, Dr. Timothy Elston, Drs. Akwari,  Drs. Haynes, Dr. John Mayer, Dr. Lex Oversteegen, Dr. Peter V. O'Neil, Dr. Anne Cusic, Dr. Henry van den Bedem, Dr. James R. Ward, Dr. Marius Nkashama, Dr. Michael W. Quick, Dr. Robert J. Lefkowitz, Dr. Arlie Petters, and several other excellent educators and role models not mentioned here.

\newpage
\section*{Appendix A}
\label{subsec:signals}

 Signals processing plays an important role in magnetic resonance imaging, and is briefly reviewed in this appendix. The free inductance decay signal is often collected in the frequency domain and Fourier transformed into the time domain~\cite{eran1966}. Also with a gradual shift towards MRI scanners of higher magnetic field, the signal-to-noise ratio requires increasingly more effective noise filtering algorithms. As signal generation and collection methods become increasingly more sophisticated e.g via spirals and other geometric sequences, non-uniform sampling, and exotic $B_1$ pulse sequence configurations, signals processing methods need to advance in tandem to address problems arising. Computing speed also takes on new significance with such developments. Indeed imaging and graphics applications have significantly driven the interest and funding for fast algorithms and computing. Notable advancements include the fast Fourier transform, the fast-multipole method of Greengard and Rokhlin~\cite{grro1987,grro1997}, and parallel computing hardware such as the Compute Unified Device Architecture (CUDA) multi-core processors by Nvidia for graphics processing~\cite{pagu2008}. Furthermore, a new field is taking form at the interface of high performance software algorithms and multi-core hardware processors~\cite{nuog2008,ma2007,nibu2008,hana2007,pari2009,keir2011}.

 We take a brief look at some of the basics of signals processing below,

\subsubsection*{A.1 Continuous and Discrete Fourier Transforms}

 The Fourier transform converts a time domain signal to its corresponding frequency domain signal, and is given by,

 \begin{equation}
  F[f(t)] = f(\omega) = \int_{-\infty}^{\infty}f(t) e^{-i\omega t}dt,
 \end{equation}

while the inverse Fourier transform converts a frequency domain signal to its corresponding time domain signal, and is given by,

\begin{equation}
 F^{-1}[f(\omega)]=f(t) = \int_{-\infty}^{\infty}f(\omega) e^{i\omega t}d\omega.
\end{equation}

In practice, the discrete versions of the above equations are used instead, and give good results provided that signal data is sampled at an appropriate frequency, the Nyquist frequency. Given an MRI signal of $N$ sample points, $f[n]$, where $n=0,..,N-1$, the frequency domain representation is given by the discrete Fourier transform,

\begin{equation}
 f[{\omega}_k] = \sum_{n = 0}^{N-1} f[n]e^{-i2\pi\frac{k}{N} n},  
\end{equation}

and the discrete inverse Fourier transform is given by,

\begin{equation}
  f[n] = \frac{1}{N}\sum_{k = 0}^{N-1}  f[{\omega}_k]e^{i2\pi\frac{k}{N} n}.
\end{equation}

The above Fourier formulas readily allow extension to two and three dimensions as well as various other customizations, tailored to suite the particular MRI application and data format.

\subsubsection*{A.2 MRI Sampling and the Aliasing Problem}

\begin{figure}[ht]
\begin{center}
\scalebox{.65}
{\includegraphics{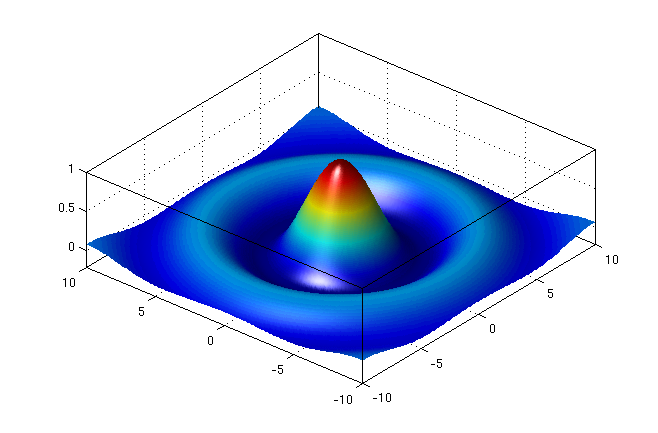}}
\end{center}
\caption{3D sinc function}
\label{fig:sinc}
\end{figure}

 The Whittaker- Kotelnikov- Shannon- Raabe- Someya- Nyquist theorem prescribes a lower bound on the sampling frequency necessary for perfect reconstruction of a band-limited signal. It states that given a band limited signal $g(t)$ whose frequency domain representation is such that $|g(f)|<B$ for all $f$ and some $B$, then by sampling at a rate $f_s=2B$, the image can be perfectly reconstructed. The sampling interval is $T = 1/f_s$, and the discrete time signal is expressed as $g[nT]$ where $n$ is an integer. $f_s={\omega}_s/2\pi$ and is in units of hertz, hence $T$ is in seconds. The perfect reconstruction is given by the Whittaker-Shannon interpolation formula,

\begin{equation}
 g(t) = \sum_{n=-\infty}^{\infty} g[nT] \cdot \mbox{sinc}\left(\frac{t-nT}{T}\right),
\end{equation}

where sinc is the \textit{normalized sampling function} given by,

\begin{equation}
 \mbox{sinc}(x) = \begin{cases}
\displaystyle 1 &~\mbox{if $x=0$},\\
\displaystyle \frac{\sin(\pi x)}{\pi x} &~\mbox{otherwise}.\\
\end{cases} 
\end{equation}

In practice, low-pass filtering methods can be used to decrease the amplitude of frequency components which exceed the effective bandlimit.

The aliasing problem arises when the sampling rate $f_s$ is less than twice the bandlimit $B$. The effect of this is that the reconstructed image is an alias of the actual image. The Discrete Time Fourier Transform (DTFT) of a signal is given by, 

\begin{equation}
 X(\omega) = \sum_{n=-\infty}^{\infty} x[n]e^{-i\omega n},
\end{equation}

and in the case of an under-sampled MRI signal, the DTFT matches that of the alias. Therefore the reconstructed image is merely an alias. This phenomenon can be described more formally by invoking the poisson summation formula,

\begin{equation}
 X_s(f):=\sum_{-\infty}^{\infty}X(f-kf_s)=\sum_{-\infty}^{\infty}T\cdot x(nT)e^{-i2\pi f nT},
\end{equation}
 
where the middle term above is the periodic summation. For a bandlimited under-sampled signal, increasing the sampling rate to satisfy the Nyquist criterion will restore perfect reconstruction.

\subsubsection*{A.3 The Convolution Theorem}

The convolution theorem facilitates application-specific image processing before or after reconstruction in either k-space or time domain. The theorem is given by,

\begin{equation}
 {\cal{F}}\{(f\ast g)(t)\} = f(\omega)\cdot g(\omega),
\end{equation}

and states that the Fourier transform of the convolution of two functions equals the product of the Fourier transforms of both functions. Where the convolution is defined as,

\begin{equation}
 (f\ast g)(t) := \int_{-\infty}^{\infty}f(\tau)g(t-\tau)d\tau.
\end{equation}

\newpage
\providecommand{\bysame}{\leavevmode\hbox to3em{\hrulefill}\thinspace}
\providecommand{\MR}{\relax\ifhmode\unskip\space\fi MR }
\providecommand{\MRhref}[2]{%
  \href{http://www.ams.org/mathscinet-getitem?mr=#1}{#2}
}
\providecommand{\href}[2]{#2}

\newpage
\section*{Author Biography}

{\small{
Dr. Stephen G. Odaibo is Chief Executive Officer and Founder of Quantum Lucid Research Laboratories. He is a Mathematician, Computer scientist, Physicist, Biochemist, and Physician. Dr. Odaibo obtained a B.S. in Mathematics (UAB, 2001), M.S. in Mathematics (UAB, 2002), M.S. in Computer Science (Duke, 2009), and Doctor of Medicine (Duke, 2010). He completed a two year pre-doctoral research fellowship in Biochemistry in affiliation with the Howard Hughes Medical Institute between 2004 and 2006, during which he studied cell signaling. Dr. Odaibo completed a housemanship in Internal Medicine at Duke University Hospital and is currently an Ophthalmology resident at Howard University Hospital in Washington DC. His awards and recognitions include: he invented the Trajectron method and provided the first quantitative demonstration of non-paraxial light bending within the human cornea; his first paper was selected by MIT Technology Review as one of the best papers from Physics or Computer science submitted to the arXiv the first week of Oct 2011; he was an Invited Discussant to the exclusive launching of the U.S. National Academy of Sciences Committee Report, \textit{Optics and Photonics: Essential Technologies for Our Nation}; he won the Barrie Hurwitz Award for Excellence in Clinical Neurology (Duke, 2005); he won the International Scholar Award for Academic Excellence (UAB, 2001); and in 2012 he was selected as a Featured Alumnus of the Mathematics Department at UAB. Dr. Odaibo's research interests are at the fusion of Mathematics, Computer Science, Physics, and the Biomedical Sciences, with a special focus on the interaction of light with biological systems. His interests also include exploring approaches towards pedagogical unification of the Natural Sciences and Biotechnology disciplines. He is happily married to Dr. Lisa Odaibo who is completing a Residency in Pediatrics.}}

\end{document}